\documentclass{article}
\pdfoutput=1
\usepackage{amsmath,amssymb}

\usepackage{latexsym,graphicx}

\usepackage[T1]{fontenc} 

\newcommand{\geom} {\rm{geom}}

\usepackage{wrapfig}

\def\lb{\linebreak[4]}
\newcommand{\be}{\begin{equation}}
\newcommand{\ee}{\end{equation}}
\newcommand{\bea}{\begin{eqnarray}}
\newcommand{\eea}{\end{eqnarray}}
\newcommand{\bes}{\begin{subequations}}
\newcommand{\ees}{\end{subequations}}

\def\ba{$$\begin{array}}
 \def\ea{\end{array}$$}
\newcommand{\beg}{\begin{equation}\begin{gathered}}
\newcommand{\eeg}{\end{gathered}\end{equation}}

\newcommand{\fr}[2]{\dfrac{{ #1}}{{ #2}}}
\newcommand{\la}{\langle}
\newcommand{\ra}{\rangle}
\newcommand{\fn}[1]{\footnote{{#1}}}
\newcommand{\bu}{$\bullet$\ }
\newcommand{\tra}{$\triangledown$ }

\renewcommand{\le}{\leqslant}
\renewcommand{\ge}{\geqslant}

\newcommand{\epe}{\mbox{$e^+e^-\,$}}
\newcommand{\ggam}{\mbox{$\gamma\gamma\,$}}
\newcommand{\egam}{\mbox{$e\gamma \,$}}
\def\cl{\centerline}

\newenvironment{Itemize}{\begin{list}{$\bullet$}%
{\setlength{\topsep}{0.2mm}\setlength{\partopsep}{0.2mm}%
\setlength{\itemsep}{0.2mm}\setlength{\parsep}{0.2mm}\setlength{\leftmargin}{4mm}}}%
{\end{list}}
\newcounter{enumct}

\begin{document}
\date{}

\title{\LARGE\bf Photon Collider for  energies 1-2 TeV}

\author{I.~F.~Ginzburg,\\
G. L. Kotkin\fn{\it Gleb Kotkin died on May 8, 2020. We, his friends, colleagues and pupils,  will remember him.},\\
\it Sobolev Institute of Mathematics, Novosibirsk, 630090, Russia;\\  Novosibirsk State University, Novosibirsk, 630090, Russia.\\
 Editorially improved version of the text published in\\ {\bf Physics of Particles and Nuclei} {\bf 52}  (2021) \#5}

\maketitle

\begin{abstract}

We discuss a photon collider based on the \epe \ linear collider with energies of $2E = 1\div 2$~TeV in cms (ILC, CLIC, ...).
Previously, this energy range was considered hopeless for the experiment in the foreseeable future. We discuss the realization of the TeV PLC based on modern lasers. A small modification of the laser-optical system prepared for the photon collider with energy of $E \approx 250$~GeV will be sufficient if the parameters are chosen optimally. The high-energy part of the photon spectrum does not depend on design details and is well separated from the low-energy part. That is a narrow band near the upper boundary, about 5\% wide. The high-energy integrated \ggam luminosity is about 1/5 and the maximum differential luminosity is about 1/4 of the corresponding values for the photon collider with $E \approx 250$~GeV.
\end{abstract}





\section{Introduction}

\bu \ {\bf Two photon  processes -- virtual  photons.}
The processes now called two-photon processes have been studied since 1934  \cite{LL}. It was the production of a \epe \ pair in a collision of ultrarelativistic charged
particles, $A_1A_2 \to A_1A_2+X$ with $X= \epe$. Next 35 years different authors considered similar processes with  $X=\mu^+\mu^-$,  or $\pi^0$, or $\pi^+\pi^-$ (point-like) (see references in review \cite{BGMS}).

In 1970, it was shown that observing the processes $A_1A_2 \to A_1A_2 + X$ at colliders (primarily at \epe colliders) will allow us to study the processes\lb $\ggam \to X$ with two quasi-real photons at very high \ggam energies $M_X$ (see \cite{BBG}, \cite{BBG1} and a little later \cite{BKT}).

The general description of such processes given in the review \cite{BGMS} is still relevant today.
The collision of particles $A_i$ with mass $M_i$,  electric charge $Z_ie$ and energy $E_i$ generates  a pair
{\it virtual} (quasi-real) photons with energies $\omega_i$.  Their fluxes (per one initial particle $A_i$) are
\beg
f(\omega_i)d\omega_i\approx\fr{Z_i\alpha}{\pi}\,g_i(\omega_i/E_i)\,\fr{d\omega_i}{\omega_i}\,L_i\theta(L_i),\quad
L_i\approx \ln \left(\lambda_q^2(E_i/M_i)^2/\omega_i^2\right),
\eeg
where the shape of the functions $g_i(x) \le 1$ depends on the type of colliding particles, and the parameter $\lambda_q$  depends on the properties of both particles $A_i$ and the produced system $X$.

The study of processes with quasi-real photons has become an important field of experiments at colliders (for example, see \cite{PHOTON17}). The results of these experiments significantly added to our knowledge of resonances
and the details of hadron physics. However, such
experiments cannot compete with experiments at other colliders when studying the problems of New
Physics. Indeed, the luminosity of collisions of virtual photons in the high-energy region is $3- 5$ orders of magnitude lower than the luminosity of current \epe or $pp$ colliders. For collisions of heavy nuclei (RHIC, LHC), the effective energy spectrum of virtual photons is bounded from above and difficulties with the signature of  \ggam events  at high photon energies are added here.

\bu \ {\bf Photon Colliders with real photons -- PLC.}
Another approach, which allows one to study  collisions of real high energy photons, was developed in 1981 when discussing the capabilities of \epe linear colliders  (\epe LC). In LC each electron bunch is used  once. Hence, one can try to convert a significant fraction of the incident electrons into photons with energies close to the electron energy, so that the collisions of these photons will compete with the main collisions both in the collision energy and its luminosity.
 The way to implement this idea -- the Photon Linear Collider (PLC) -- was proposed in \cite{GKST1} -- \cite{GKST3}.

\begin{wrapfigure}[12]{l}[0pt]{6.5cm}
\includegraphics[width=0.45\textwidth]{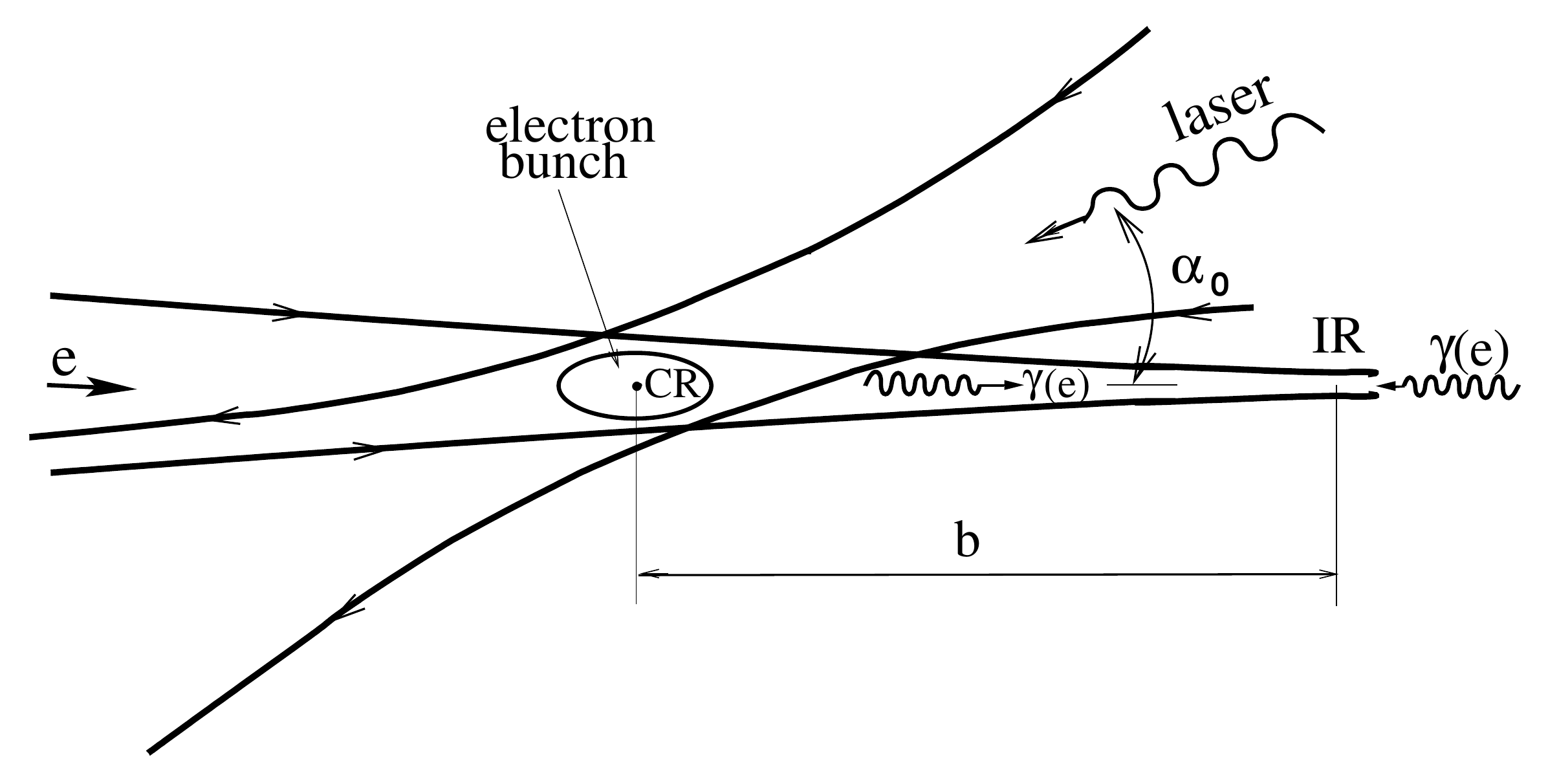}
\caption{\it Scheme of PLC } \label{fig:basschem}
\end{wrapfigure}
The well-known scheme of \  PLC
is shown in Fig.~\ref{fig:basschem}. In the conversion  region
 $CR$ preceding the  interaction  region $IR$,
 electron ($ e^-$ or $ e^+$) beam of the  LC
 encounters a photon beam (a flash of a powerful laser).
The Compton scattering of laser photon on electron from LC (with energy $E$) generates a photon  with energy close to $E$,\fn{\it For definiteness, we discuss the case when both initial
beams are electron beams $e^-$.

We distinguish electron in the collider bunch $e_o$ and produced electron $e$, photon in the laser flash $\gamma_o$ and produced  photon $\gamma$.}
\be
e_o\gamma_o\to e\gamma\,.\label{basComp}
\ee
 These photons
are focused in the IR  into a spot of about the same size as expected for electrons without laser conversion.
In the IR these photons
collide with photons from the opposite conversion region
(\ggam collisions) or with electrons from  counter-propagating beam
 ($\gamma e$ collisions).

The ratio of the number of high energy photons to that of electrons in LC is called
the conversion coefficient $k$, for the standard PLC $k\approx
1-1/e\approx 0.63$ typically.

Numerous studies of PLC (see, for example,
\cite{Tel90}, \cite{TESLA}, \cite{Tel16}) are developing many new technical details,\fn{\it A recent example is optimization of the beam crossing angle  for \epe and \ggam  collisions  at the LC \cite{Tel18}.} but
they all retain the original scheme of Fig.~\ref{fig:basschem}.

The main properties of the basic Compton process are determined by the  parameter
 \be
 x=4E\omega_0/m_e^2\,,
\label{xdef}
\ee
where $E$ is the electron beam energy and
$\omega_0$ --  the laser photon energy. (To simplify
text, we  set $\alpha_0=0$ in Fig.~\ref{fig:basschem}.)

 In 1981, to construct PLC, it was proposed to use a laser with neodymium glass or garnet, for which $\omega_o = 1.17$~eV \cite {GKST1}-\cite {GKST2}.  For electrons with energy $E\le 250$~GeV,
 this  choice remains optimal until now. Using such a laser, we have
 \beg\label{EXval}
2E=0.5\; TeV\;\Rightarrow\; x=4.5;\\
2E=1\; TeV\;\Rightarrow\; x= 9;\\
2E=2\; TeV\;\Rightarrow\; x= 18;\\
2E=11\; TeV\Rightarrow\; x= 100.
\eeg
First three lines here correspond different stages of ILC and CLIC projects, the fourth line represents some aymptotics.

\bu \ The described scheme in its pure form works only at $x <2 (1+ \sqrt {2}) = 4.8$ (at
$E <270$~GeV for mentioned laser). At higher values of $x$, which will be realized at the subsequent stages of ILC, CLIC,..., some of the resulting high-energy photons die out, forming the \epe pairs in collisions with laser photons from the tail of a laser flash,
\be
\gamma\gamma_o\to \epe\quad (killing\;\; process) \,.\label{killing}
\ee
This fact was considered as limiting the possibility of implementing a PLC based on
LC with high electron energies
 \cite{GKST1}, \cite{GKST2}).

Two ways to overcome this difficulty are discussed.\\
(1) To use a new laser with lower photon energy, keeping $x\lesssim 4.8$;\\
(2) To use existing laser, resigned to the decrease
in \ggam \ luminosity.

 \tra  The results of \cite{GKST1} -- \cite{GKST3} are directly applicable for the \underline{first way}.
However, a new laser is required for each new energy of electrons; so far, such lasers with the required parameters have not been developed.

\tra , This work is devoted to the study of \underline{the second way} with the guiding idea to get almost monochromatic collisions of photons at a cost an acceptable decrease in luminosity. This possibility is noted in \cite{Tel2001,GKPho9} and partially developed in \cite{Tel2001}.

Unfortunately, the method of optimizing the conditions for conversion $e\to\gamma$ used in \cite{Tel2001} gives an inaccurate result for some beam polarizations. The analysis of the spectra of final photons without considering the polarization of particles distorts the spectra and the luminosity distribution. The result is obtained only for a certain configuration of the facility. It is not clear what changes should be expected when the engineering solution for beam collision changes. We correct these inaccuracies below and find out to what extent the results can be applied to any engineering solution for beam collisions. {\it We call such photon collider the TeV PLC.}

\bu \ {\bf Desirability of magnetic deflection of the electrons along the CR--IR path.}
In the papers \cite{GKST1}-\cite{GKST3} it was proposed to remove electrons that survived after conversion from the interaction region using a transverse magnetic field along the CR - IR path.
For $ x<4.8$, the problem was to obtain the maximum conversion coefficient so that after passing through the CR, only a small part of the electrons retained their initial energy, and   most of electrons decrease energy and scatter away at small angles, similar to photons. This effect  increased the total spread of electrons in the transverse plane. In addition, when using colliding $e ^-e ^ - $ beams, residual electrons in the IR are repulsed by Coulomb forces so that the interaction of these electrons with each other and with the resulting photon beams becomes insignificant. Therefore, it is possible to dispense with the inclusion of this  magnetic field
\cite{Tel90}. The PLC projects discussed in \cite{TESLA}-\cite{Tel18} do not
have such a magnetic field.

For $x>4.8$, in the optimal situation (see below), about one half of the electrons pass through the CR freely (without interaction with photons). In the absence of a magnetic field, it would be wrong to neglect their interaction with counterpropagating photons and electrons. To eliminate this interaction, electrons must be removed from the interaction region using a transverse magnetic field along the CR-IR path, as it was suggested in \cite{GKST1}-\cite{GKST3}. Further, we assume {\bf for $\pmb{ x> 4.8} $ that when studying \ggam \ collisions the magnetic field is turned on for both beams, and when studying \egam \ collisions the magnetic field is turned on for one beam (converted to photon).}\label{pagemagn}\\

\bu \ {\it\Large The organization of the subsequent text is as follows.}

Section \ref{sectechnintro} contains basic notation and a description of the Compton process in the studied parameter range with examples for $x =4.5$. We also introduce the important concept of the optical length of a laser flash for high-energy electrons.

In Section \ref{secphotinIR}, we discuss the sources of photons falling into the interaction region. It is shown that the energy spectrum of these photons and the corresponding luminosities are naturally divided into two parts that are well separated from each other. The high- energy part, which is most interesting for studying the problems of New Physics, admits a universal description independent of the details of the experimental facility.
There is no such universal description for the
low-energy part. The basic notations related to the distribution of high-energy luminosity are also introduced in this section.

In the main part of the work  (Section~\ref{secbigx}), we discuss the case  $x > 4.8$ and the main characteristics of high-energy \ggam and \egam collisions considering the modifications introduced by new processes in CR. We consider the cases $x=9,\;18$ (ILC, CLIC) in detail and give some
examples for $x=100$.  After a brief
discussion of the basic Compton effect at $x>4.8$ (Section~\ref{secbassp}),
Section~\ref{seckilling} discusses the killing process $\gamma\gamma_o\to\epe$, which was not discussed in necessary detail earlier, because it takes  place only for $x > 4.8$. In section~\ref{seceq}
we write down the balance equations for the
number of photons, produced in the Compton process and lost within the conversion region due to the killing process. The polarizing properties of the killing process turned out to be very significant.

The next step is to choose the optimal value for the laser flash energy or, in other words, the optical length of the laser flash.  We have chosen as a criterion to obtain the maximum number of collided photons for the physics problems of interest (Section~\ref{secopt}).

Section~\ref{secdistr} contains a description of the resulting high-energy spectra for $\egam$ and $\ggam$ collisions.

A brief description of the results is given in Section~\ref{secsum}.\\

In appendix~\ref{secbad} we discuss the case of ``bad'' choice of initial polarization.
In appendix ~ \ref {seclinear} we find out that research with linear polarization of a high-energy photon is practically impracticable at TeV PLC.
The most important background is discussed in appendix~\ref{appa}.
In appendix ~ \ref{BHproceq} we discuss the Bethe--Heitler process that could reduce the number of produced photons and show that this phenomenon is insignificant up to very high energies.

In appendix~\ref{secphys}, we list some of the important New Physics problems that can be explored using TeV PLC and cannot be studied using the LHC and \epe colliders

\section{Some notations, Compton effect on a high-energy electron}\label{sectechnintro}

For definiteness, we consider as a base the $e^-e^-$ LC (not
 $e^+e^-$). We neglect the effects of high photon density in the conversion region (nonlinear QED effects). In all discussions, we assume $E\gg\omega_o$.

{\bf 1. Notations.}\label{pagenot}
\begin{Itemize}
\item $\gamma_o$, $\omega_o$, $\lambda_o$ --   the  laser photon,  its energy and helicity;

\item $e_o$, $E $, $\lambda_e$ --  the  initial electron, its energy and helicity ($2|\lambda_e|\le 1$);

\item  $\gamma$, $\omega$, $\lambda$ --  the  produced photon, its energy and helicity;
\item $\sigma_0=\pi r_e^2=\;2.5\cdot 10^{-25}\;\mbox {cm} ^{2}$;

\item $b$ --  the  distance  between  the conversion  region $CR$ and  the
interaction region $IR$ (Fig.~\ref{fig:basschem});

\item   $\Lambda_C=2\lambda_e\lambda_o$ --  polarization parameter  of   the process;
\item $y=\omega/E$  --  relative photon energy;
\item $y_M=x/(x+1)$ --  maximal value of $y$ at given $x$;
\clearpage
\item $y_{\min}$  --- position of the minimum in the dependence $d\sigma/dy$ on $y$ for the Compton process at $\Lambda_C\approx -1$ (defined for $x\ge 3$, it depends on $x$)\fn{\it  For  $\Lambda_C<-0.75$ it is found that ({\it i})
 the total fraction of photons with $y>y_{min}$ is more than 0.55 Eq.~\eqref{taudef},  ({\it ii}) it is weakly depends on $\Lambda_C$. For calculations, it is more convenient to define $y_{min}$ via Eq.
 \eqref{taudef}.
};
\item
$w_{\gamma\gamma}=\fr{\sqrt{4\omega_1\omega_2}}{2E}\equiv \sqrt{y_1y_2}\le y_M$,\\
 $w_{e\gamma}=\fr{\sqrt{4E\omega_1}}{2E}\equiv \sqrt{y_1}\le \sqrt{y_M}$ --
ratios
of the \ggam \  and \egam \ cms energies to $\sqrt{s}=2E$.

\end{Itemize}

{\bf 2. Compton scattering.}
We present the basic information about the Compton backward scattering  in the kinematic
region of interest according \cite{GKST1} -- \cite{GKST3}  in the form, suitable for description of high energy
part of spectra  at large $x$,  with addition of some details which were not discussed earlier.
Here  numerical examples are given for  $x=4.5$, which is close   to   the upper limit
of validity  in previous  studies.

The total cross section of Compton effect is well known (Table~\ref{tabCompxsec}):
\beg
\sigma_C=\fr{2\sigma_0}{x}\left[F(x)+\Lambda_C T(x)\right], \\
F(x)\!=\!\left(1-\fr{4}{ x}\! -\! \fr{8}{x^2}\right)\ln (x+1)\! +\!\fr{1}{2} \!+\!\fr{8}{x}\! -\! \fr{1}{ 2(x+1)^{2}} ,
 \\
T(x)\!=\!\left(1\!+\!\fr{2}{x}\right) \ln (x+1)\!-\!\fr{5}{2}\!+\!\fr{1}{ x+1}\!-\!\fr{1}{2(x+1)^{2}}.
  \label{xComp}\eeg

  \begin{table}[hbt]\label{tabCompxsec}\begin{center}

\begin{tabular}{|c||c|c| c|c|}\hline
$x$&4.5&9&18&100\\\hline
$\Lambda_C\!=\!-1$&0.73&0.45&0.26&0.056\\\hline
$\Lambda_C=1$&0.85&0.63&0.44&0.145\\\hline
\end{tabular}
\caption{\it
Compton cross section $\sigma_C/\sigma_0$ at  different $x$ and $\Lambda_C$}\label{tabCompxsec}\end{center}
\end{table}
The photon energy is
kinematically bounded from above by quantity $y_M=x/(x+1)$ ($y_M=0.82$ for $x=4.5$).
The  energy distribution of
photons strongly depends on $x$ and $\Lambda_C$ (here $ r=\fr{y}{x(1-y)}\le 1$)
 \beg
f(x,y)\!\equiv \!\fr{1}{\sigma_
 C}\, \fr{d\sigma _C}{
dy}=\fr{U(x,y)}{F(x)+\Lambda_C T(x)}\,,\\
U(x,y)=\fr{1}{1-y} + 1-y - 4r(1-r)-
\Lambda_C x r(2-y)(2r-1). \label{yComp}
 \eeg

\begin{wrapfigure}[14]{l}[0pt]{5.5cm}\vspace{-5mm}
\includegraphics[width=0.45\textwidth]{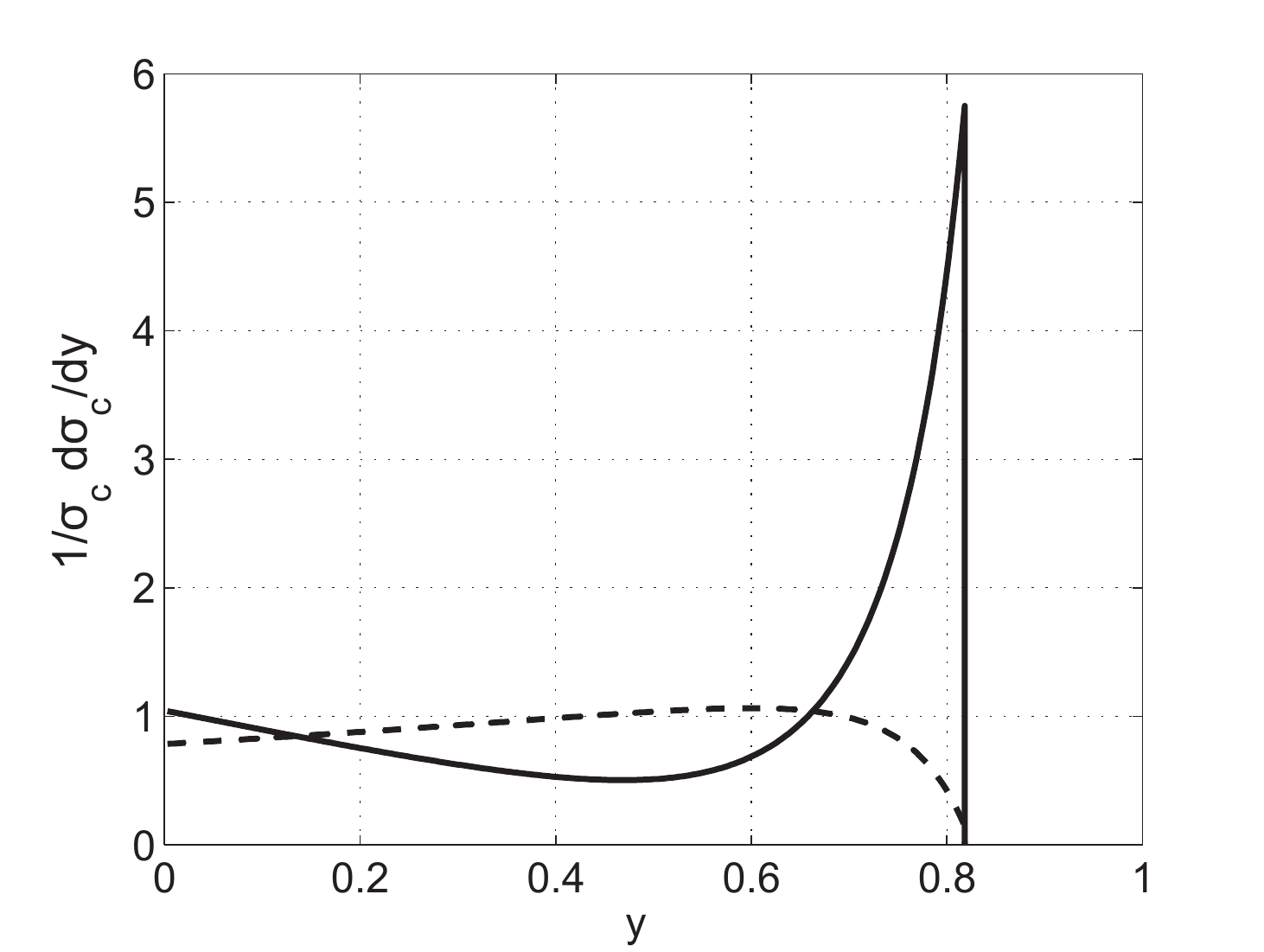}
\vspace{-6mm}
\caption{\it Photon energy spectrum $f(x,y)$ at $x=4.5$, $\Lambda_C=-1$ (solid)
and $\Lambda_C=1$ (dotted)} \label{fig:Compsp48}
\end{wrapfigure}

 At $\Lambda_C\approx-1$,  the
photon spectrum grows up to its upper boundary $y=y_M$, at
$\Lambda_C\approx 1$ this spectrum is much  flatter
(Figure~\ref{fig:Compsp48}).

 At $x>3$ and $\Lambda_C\approx -1$ the high
energy part of this distribution is concentrated in the narrow band below upper limit,  contained more than one half produced photons. We  characterize this band by its lower boundary
$y_{\min}$ and the sharpness parameter $\tau$ -- see Table~\ref{hepeak}
\beg
\!\!\mbox{\it alternative definition of $y_{\min}$:} \!\!\int\limits_{y_{\min}}^{y_M}\!\!f(x,y)dy\approx 0.55;\\
\mbox{\it definition of $\tau$:} f(x,y=y_M(1-\tau))\!=\!\fr{1}{2}\cdot f(x,y_M).\label{taudef}
\eeg
\begin{table}[h]
\centering{
\begin{tabular}{|c||c|c|c|c|}\hline
x& 4.5&9&18&100\\\hline\hline
$y_{\min}$ &0.6&0.7&0.75&0.94\\\hline
$\tau$&0.09&0.036&0.022&0.004\\\hline
\end{tabular}
\caption{\it Properties of high energy peak at $\Lambda_C\approx -1$}
\label{hepeak}}
\end{table}

The mean circular polarization  of   the produced photon (helicity) is
\bes \label{polComp}\beg
 \lambda(y) =\\=\fr{
\lambda_o(1\!-\!2r)\left[\fr{1}{1\!-\!y}\!+\!1\!-\!y\right]\!+\!2\lambda_e
xr[1\!+\!(1\!-\!y)(2r\!-\!1)^2]}{U(x,y)}.\label{polComp1}
  \eeg
At $\lambda_o=\pm 1$, this equation is simplified:
\be
 \!\!\lambda(y)
\!=\!-\!\lambda_o\left[(2r\!-\!\!1)\!+\!\fr{
\left[(2r\!-\!1)\!-\!\Lambda_C xr\right]4r(1-r)} {U(x,y)}\right]\!.\label{polComp2}
  \ee
  \ees
Maximum energy photons   are well polarized with the same
direction of spin as laser photons, i.~e. $ \lambda(y=y_M)
=-\lambda_o$.

These photons
  move in the direction of the initial electron.
The emission angle of a photon increases as its energy decreases
 \be
\theta (y)=\fr{m }{E}\; \sqrt{x+1}\; \sqrt{\fr{y_M}{ y}-1}.
 \label{thety}
 \ee

Due to the increase in the angular spread on the
way from the conversion region CR to the interaction region IR, softer photons are distributed over a wider region and collide less frequently than hard photons; their relative contribution to the luminosity decreases. As a result, the high-energy part of the resulting photon spectrum is better and better separated from the low-energy part as the distance $b$ increases together with decreasing total luminosity; this separation is enhanced with increasing energy $E$ of the colliding electrons.

{\bf 3. The optical length of   the laser bunch.}
The laser flash is assumed to be wide enough so that the
inhomogeneity of the electron density inside the electron bunch is insignificant

The optical length of   the laser bunch for electrons is expressed
via   the  longitudinal density of photons  in flash  (i.~e. via the laser bunch energy $A$ divided to its effective transverse cross section
 $S_L$) and  the total
cross section of   the Compton scattering $\sigma_C(x,
\Lambda_C)$:
\be
z =\fr{A}{\omega_o S_L}\,\sigma_C(x,
\Lambda_C)\equiv\fr{A}{A_0}
\fr{\sigma_C(x, \Lambda_C)}{\sigma_C(x\!=\!4.5, \Lambda_C\!=\!-1)}\,.
\label{ZAcorresp}
\ee
In the second form of this definition we introduce $A_0$ -- the laser flash energy, necessary to
obtain $z=1$ at $x=4.5$, $\Lambda_C=-1$.

When  the electrons traverse
 the laser beam, their number decreases as
\be
n_e(z)=n_{eo}e^{-z}\,.\label{numbel}
\ee

\section{Photons in the IR}\label{secphotinIR}

Let us list the sources of photons in the interaction region.

The highest energies have photons produced in the main Compton effect. One can see that the number of
such photons with energies greater than $y_{min}E$ is about 15\% of the number of initial electrons after the loss of
some of the photons in the killing process for $\Lambda_C\approx -1$
and for moderate distances from the conversion region
to the interaction region under optimal conversion conditions (see below).

In addition, a significant number of photons of a different origin cross the interaction region.

(A)	Photons produced from the scattering of the
tail of the laser pulse  (i) on electrons, which slow down
in the main Compton effect (rescattering) or (ii) on positrons (electrons) produced in the killing process.
These photons primarily had lower energies, and their angular distribution is wider than that of the initial Compton photons.

(B)	Photons produced from magnetic lenses focusing beams and from synchrotron radiation with magnetic deflection of beams along the CR-IR beam
path. Almost all of these photons have energies lower than $y_{min}E$.

(C)	Photons produced from the interaction of electrons with each other, which survived in conversion
(beamstrahlung). When using magnetic deflection, the number of such photons with energy greater than $y_{min}E$ is small.

\bu \ 	{\it Two regions in the energy distribution.} Thus, the luminosity spectrum of \ggam collision is naturally divided into two regions: (i) a high-energy region for $w>y_{min}E$
and (ii) a low-energy region for lower values $w$. These regions are well separated from each other.

The magnitude and shape of the low-energy luminosity obviously depend on the details of the experimental facility. They are not discussed in this work.

\bu \ {\bf High-energy  luminosity and  selection of events.}
For the problems of New Physics high-energy photons with $\omega\sim E$ are important. We study the high-energy luminosity only, considering the selection of events, which include the
states with total energy\fn{\it
The limiting value $E_{lim}$ should be determined more accurately
when simulating a real facility. Note that the total energy of the
\ggam collision products can be close to $2E$. According to our esti-
mates, the value $E_{lim}$ may turn out to be even less than E in
some cases.} $\varepsilon>E_{lim}\sim E$.
This part of the luminosity spectrum is formed by photons produced in the main Compton effect \eqref{basComp}, some of them disappear in the killing process \eqref{killing}.\label{pageselec}

\bu \ {\bf Luminosities.} We consider relative luminosities  $\cal L$, and  high energy luminosity integral
\beg
L_{\rm exp}(w)={\cal L}(w)\cdot L_{\rm geom},\\ {\cal L}_{\rm h.e.}   =\!\!\int\limits_{y_{\min}}^1 \!\!{\cal L}(w)dw\;\;\left(w=w_{e\gamma} \;\; or \;\; w_{\gamma\gamma}\right).\label{Lumintdef}
\eeg

Here $L_{\rm geom}$ is  the luminosity of the $e^-e^-$ collider
designed for \ggam \ mode.\fn{\it In  \epe LC, radiation from collisions of $e^+$ and $e^-$ bunches in
IR (beamstrahlung) limits the densities of colliding beams. In PLC, there is no such problem. This allows to have $L_{geom}$  to
be larger than the expected luminosity of the conventional \epe collider.} In a nominal ILC option, i.e. at the  electron beam
energy of 250~GeV, the geometric luminosity  can reach  $L_{\rm geom}=12\cdot 10^{34}$ cm$^{-2}$s$^{-1}$ which is
about 4 times greater then the anticipated \epe  luminosity.

All subsequent calculations are performed for the case of a "good" polarization of collided electrons and photons $\Lambda_C\approx -1$  {\bf for high energy part of luminosity},
in two versions: $\Lambda_C= -1$ and $\Lambda_C= -0.86$.

We discuss luminosities $L_I$ for different values of the total helicity
$I=|\lambda_1-\lambda_2|$ of initial state  (these are $L_{1/2}$ and $L_{3/2}$ for \egam collisions, and ($L_0$ and $L_2$ for \ggam \ collisions).
We found that one of these helicities dominates in the high energy part of the luminosity spectrum.
Therefore, below we discuss
total luminosity (sum over both finite helicities) and
indicate the contribution to it from states with non-leading complete helicity.

In addition to notations, listed in the Section~\ref{sectechnintro}, we  also
define quantities that depend on $x$, $\Lambda_C$ and $b$:
\\
\bu  ${\cal L}_{h.e.}$ --- total luminosity of the high energy peak \eqref{Lumintdef};\\
 \bu  $L_m= L(w_M)$ -- the maximal value of $L(w)$;\\
 \bu $w_M$ --  position of maximum
 in the dependence $L(w)$;\\
\bu   $w_\pm$ -- solutions of equation  $L(w_\pm)=L_m/2$;\\
\bu  $\gamma_w=(w_+-w_-)/w_M$ --  relative width of obtained peak;\\
(For \egam \ mode  $w_+=w_m=\sqrt{y_M}$,
$L_m$ don't depend on the distance $b$,\lb
$\gamma_w=1-w_-/w_M$.)\label{pagewdef}

Luminosities of \ggam and \egam collisions are given by the convolution of the spectrum of a high-energy photon with the spectrum of a counter-propagating photon or electron, considering in the calculations transverse widening of the photon beam on the way from the conversion region CR to the interaction region IR. Real beams of electrons in LC are elliptical in the transverse direction; we denote by $\sigma_x$ and $\sigma_y$ the semi-axes of the ellipse in the interaction region. This ellipticity does not allow to use  formulas
from  \cite{GKST1}--\cite{GKST3}, and for each discussed CR--IR configuration, a new simulation is usually performed (for example, see \cite{Tel90}--\cite{Tel18}).

\bu \ {\bf Simple parameterization.} In \cite{GKrho}, we found that (only at $y > y_{min}$) the effect of beam broadening on the way  CR--IR is described with good accuracy by a single parameter (as for a round beam)
 \be
\rho^2=\left(\fr{b}{(E/m_e)\sigma_x}\right)^2+\left(\fr{b}{(E/m_e)\sigma_y}\right)^2\,.
\label{rhodef}
 \ee
In other words, the shape of the high-energy part of the luminosity distribution is determined in an universal way regardless of the details of experimental facility.

In this approximation,  the discussed luminosities
are expressed via distributions in the photon energy by formulas, where
$\phi_a=\sqrt{y_M/y_a-1}$ and $I_0(z)$ is  the modified Bessel function,
 \cite{GKST1} -- \cite{GKST3}:
\bea
&\fr{dL_{\gamma e}^{\lambda}}{dy}=n_e n_\lambda (y,z)
e^{-\fr{\rho^2\phi^2}{2}}\,;
 &\\\label{Legam}
&\!\!\!\!\fr{dL_{\gamma\gamma}^{2,\lambda_1\lambda_2}}{dy_1dy_2}\!=\!
n_{\lambda_1} (y_1,z)n_{\lambda_2} (y_2,z) I_0
(\rho^2\phi_1\phi_2)e^{-\,\fr{\rho^2(\phi_1^2+\phi_2^2)}{2}}\,.&
 \label{Lggam}
 \eea
The distributions over the center of mass energy $w$ are obtained by the
substitution $w=\sqrt{y}$ \ for \egam \ luminosity or
$w=\sqrt{y_1y_2}$ with simple integration, for \ggam \ luminosity.

At $x\le 4.8$ each lost electron produces the photon.
Therefore at $z=1$ and $\rho=0$ \  total \egam \ and \ggam \ luminosities are
\be
{\cal L}_{e\gamma} =(1-1/e)=0.63\,,\quad
{\cal L}_{\gamma\gamma} =(1-1/e)^2=0.4.\label{lumz1}
\ee

\begin{figure}[h!]
\centering{\includegraphics[width=0.45\textwidth,height=0.15\textheight]{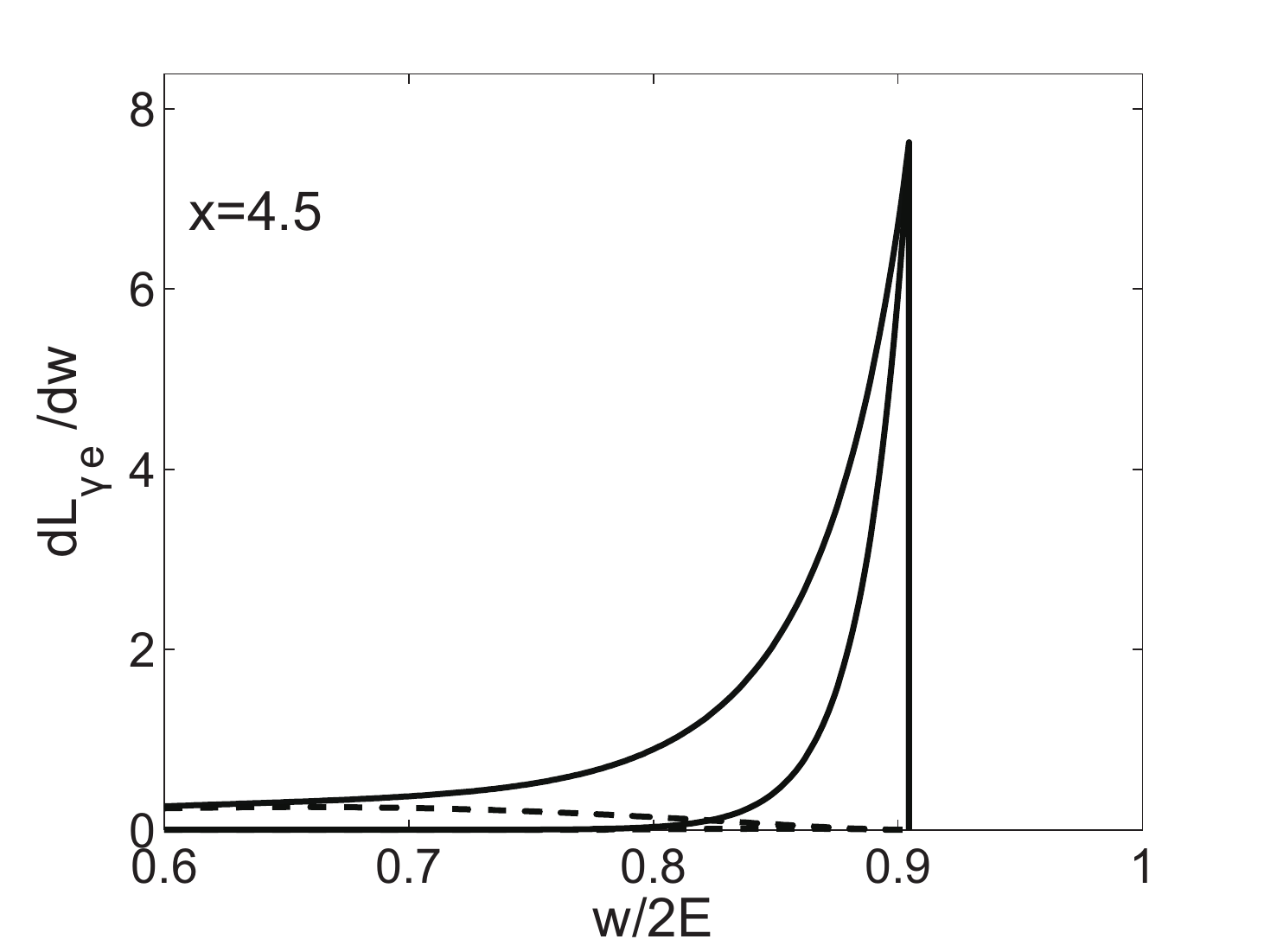}\hspace{5mm}
\includegraphics[width=0.45\textwidth,height=0.15\textheight]{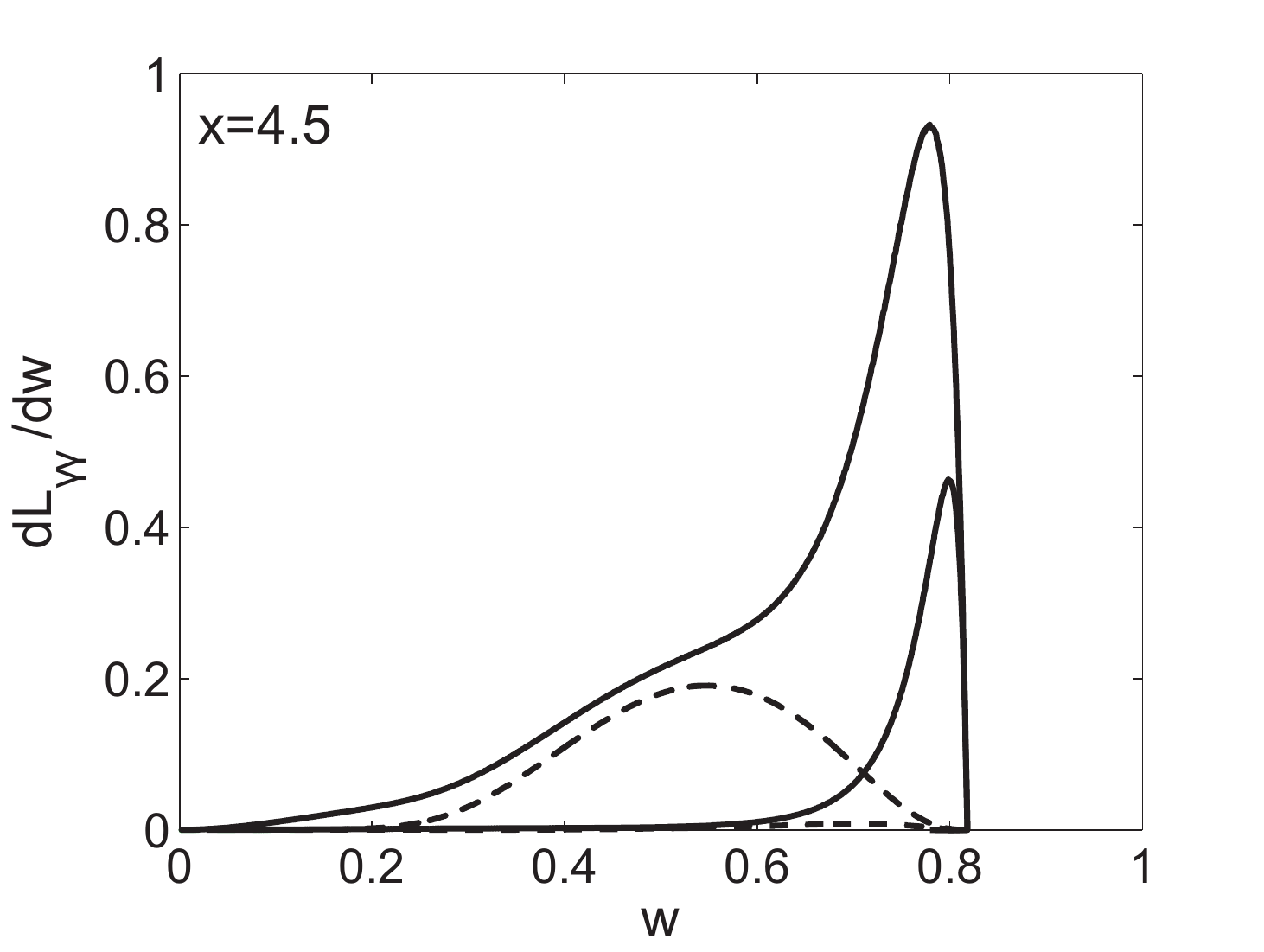}}  
\caption{\it Luminosity spectra $dL/dw$ at $x=4.5$, $\Lambda_C=-1$, $z=1$  for $\rho=1$ (up) and $\rho=5$ (down) -- solid lines, dotted lines -- similar distributions at $\rho=1$ for final states with total helicity $3/2$  for \egam \ collisions (left figure) and total helicity 2 for \ggam \ collisions (right figure).}
  \label{fig:lumsp45}
\end{figure}

 Fig.~\ref{fig:lumsp45} and first lines of
Tables~\ref{tabggamlumLam1}, \ref{tabegamlum45918} represent \ggam
\   and \ \egam \ luminosity distributions in their cms  energy
at  $z=1$, $\Lambda_C=-1$ for $\rho=1$ and 5.

\section{What happens at $\pmb{x\ge 4.8}$}\label{secbigx}

\subsection{Basic spectra}\label{secbassp}

\begin{figure}[hb]
\begin{center}
\includegraphics[width=0.4\textwidth,height=0.15\textheight]{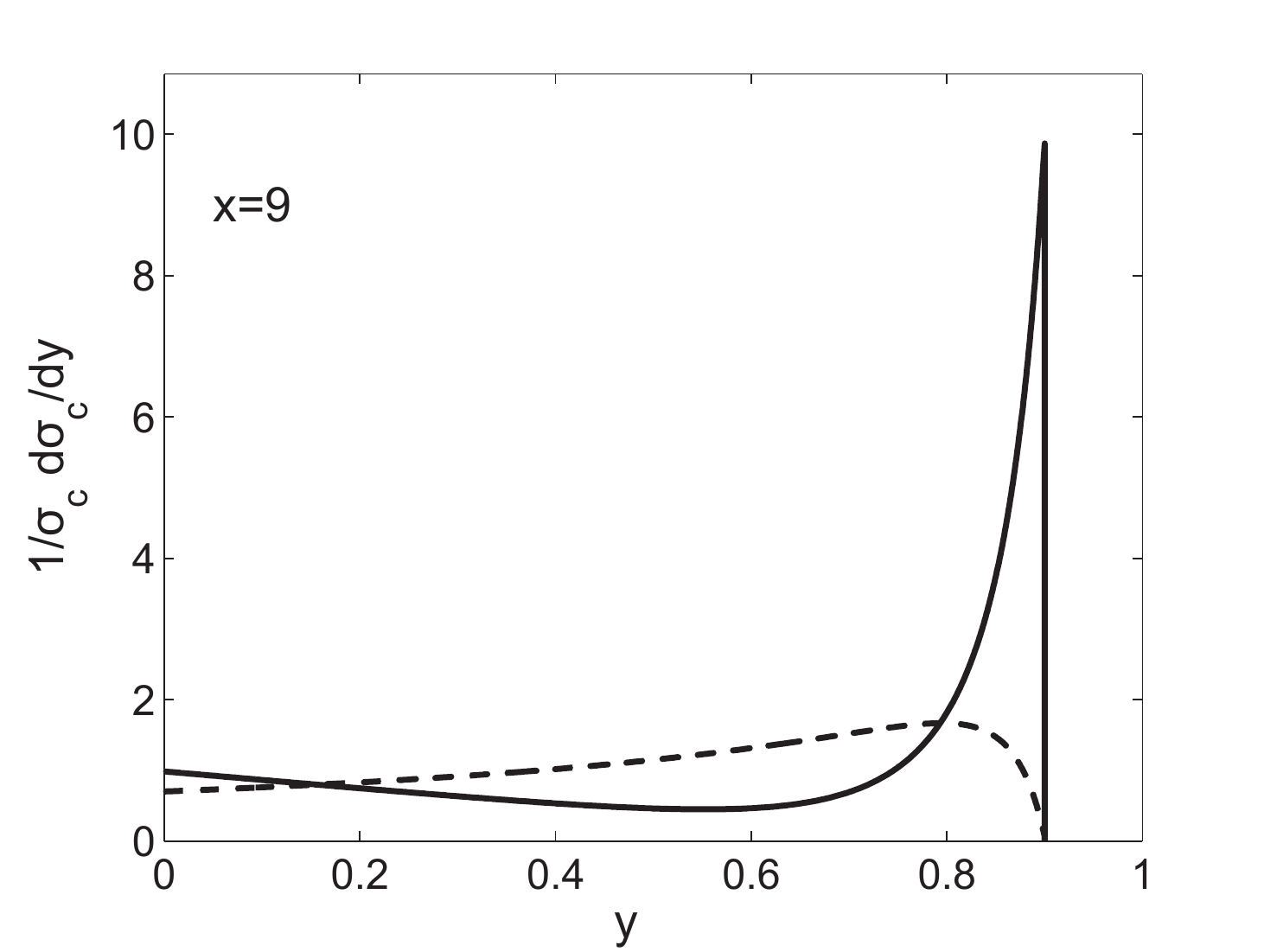}
\includegraphics[width=0.4\textwidth,height=0.15\textheight]{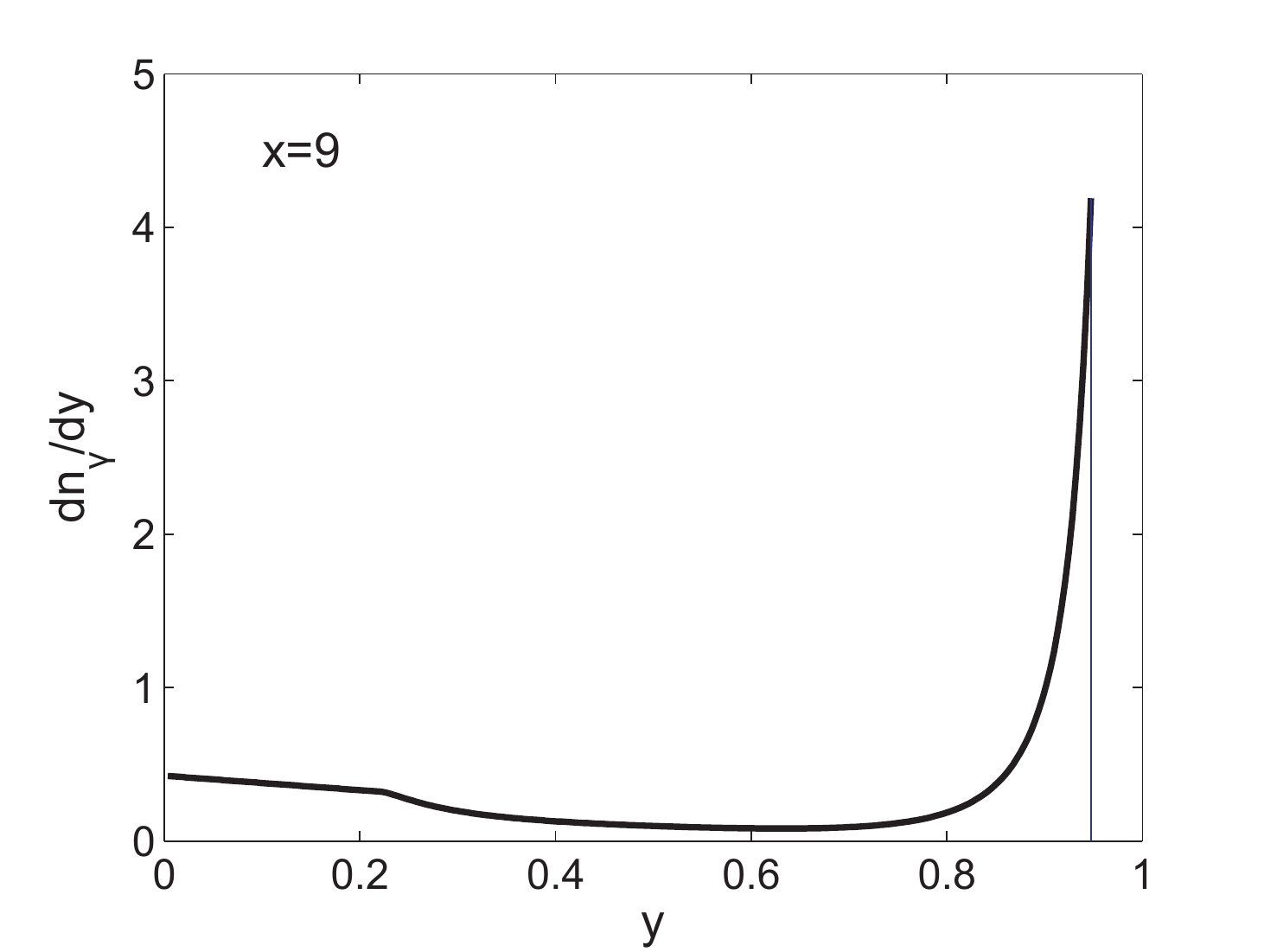}
\label{x918spectra}
\end{center}
\caption{\it The case $x=9$.
Left panel -- Compton spectrum of photons.
Solid line corresponds  to  $\Lambda_C=-1$,
dashed line --  to $\Lambda_C=1$.
Right panel --
photon energy spectrum  reduced by
killing process at $\Lambda_C=-1$, $z=0.7$}
\end{figure}

The photon energy spectrum for the basic Compton
backscattering \eqref{yComp}  at $x=9$  is shown  in
Fig.~\ref{x918spectra} -- top panel. It can be seen that this spectrum for $\Lambda_C=-1$ is
concentrated near the high energy limit  strongly then the
corresponding spectrum for  $x=4.5$ in
Fig.~\ref{fig:Compsp48}.
At $x=18$ similar calculations reveals that the
spectrum is concentrated in the   band $0.85<y\equiv
\omega/E <0.95$.

At $\Lambda_C=1$  spectrum  is almost flat, with a minimum instead of a peak near the upper boundary.

\subsection{Killing process {\boldmath $\gamma\gamma_o\to \epe$}}\label{seckilling}

The killing process is responsible for the disappearance of the Compton high-energy photons in their collisions with  laser photons from the tail of a bunch, generating
\epe \ pairs, this process is switches on at  $x\approx 4.8$.

 For a photon with energy $yE$ the squared cms  energy for killing process is $w^2_k=4\omega\omega_o/m_e^2=xy>4$. Its cross section is
\beg
\sigma_{\rm kill}(w^2_k,\lambda_o\lambda)=\fr{4\sigma_0}{w^2}\Phi_{\ggam}(w^2_k,\lambda_o\lambda)\,,\\ \Phi_{\ggam}(w^2_k,\lambda_o\lambda)=
\left(1+\fr{4}{w^2_k}-\fr{8}{w^4_k}\right)L-\\-
\left(1+\fr{4}{w^2_k}\right)v-
\lambda_o\lambda(L\!-\!3v)
,\\
v=\sqrt{1-4/w^2_k\;},\quad L=2\ln\left(\fr{w_k}{2}(1+v)\right)\,.
\label{killproc}\eeg
Note that  $\sigma_{\rm kill}(\lambda_o\lambda>0)<\sigma_{\rm kill}(\lambda_o\lambda<0)$  at $w^2_k<15$.
This inequality changes sign at $w^2_k>15$.

\subsection{Equations}\label{seceq}

The balance of  the number of high energy photons is given by their
production in the Compton process \eqref{basComp} and their
disappearance at the production of \epe \ pairs   in the
killing process \eqref{killing}.

Let us denote by $n_\gamma(y,z,\lambda)$ the flux of photons (per
1 electron) with  the energy $yE$ and polarization $\lambda$
after travelling  inside the laser beam with  the
optical length $z$. For calculations, it is convenient to split this  flux into
the sum of fluxes of right polarized photons $n_{(\gamma +)}(y,z)$
and left polarized photons $n_{(\gamma -)}(y,z)$
so  that the total photon flux $n_\gamma(y,z)$ and its average
polarization $\lambda$ are written as
 \beg
 n_\gamma(y,z)=n_{(\gamma +)}(y,z)+n_{(\gamma -)}(y,z), \\ \la\lambda(y,z)\ra=\fr{n_{(\gamma +)}(y,z)-n_{(\gamma -)}(y,z)}{n_{(\gamma +)}(y,z)+n_{(\gamma -)}(y,z)}\,.\label{densdef}
\eeg
Naturally,  $\la\lambda(y, z\to 0)\ra\to\lambda(y)$ from \eqref{polComp}.

The variation of in these fluxes during the passage
through a laser beam is described by the equations
\beg
\fr{dn_{(\gamma\pm)}(y,z)}{dz}=\\=
\!\fr{1}{2}\left(1\pm
\lambda(y)\right) f(x,y)n_e(z) -
n_{(\gamma\pm)}\fr{\sigma_{\rm kill}(xy,
\pm\lambda_o)}{\sigma_C(x)}.\label{eqdens}
 \eeg
The number of photons and their mean polarization are expressed
via auxiliary quantities $\nu_\pm$:
 \ba{c}
n_{\gamma\pm}(y,z)=f(x,y)\nu_\pm(z,y),\quad
\la\lambda\ra=\fr{\nu_+(z,y)-\nu_-(z,y)}{\nu_+(z,y)+\nu_-(z,y)}.
\ea
The equation \eqref{eqdens} is easily solved:
 \beg
n_{(\gamma\pm)}(y,z)=f(x,y)n_{e0}\nu_\pm(y,z);\\ \mbox{where}\;\;
\nu_\pm(y,z)=
\fr{\left(1\pm \Lambda_C(y)\right)}{2}
\cdot\fr{e^{-\zeta_\pm z}- e^{-z}}{1-\zeta_\pm};\\
\zeta_\pm= \fr{\sigma_{\rm kill}(xy, \pm\lambda_o)}{\sigma_C(x)} \,.
\label{solveq}\eeg

It is useful for future discussions to define in addition ratio of number of killed photons to the number of  photons, prepared for the \ggam \ collisions, $r_K(y,z)$ --  Table~\ref{1bbb}.

In Fig. 4, we compare the energy spectrum of
Compton photons (top panel) with what remains after
the passage of a laser beam with an optical length $z= 0.7$
(bottom panel).  One
can see   that

({\it i}) The shape of high-energy part of the spectrum is very similar
to that for the pure Compton effect. (This similarity
is due to the special relationship between the spin structures
of processes (2) and (5). Neglecting this structure for process
(5), as it was done in [22], one arrives at an erroneous conclusion
about a strong <<sharpening>> of the resulting spectrum.)

 ({\it ii}) The killing process <<eats>> away photons from
the middle part of the energy spectrum (improving the
separation of the high-energy and low-energy parts of
the spectrum).

({\it iii}) The separation of the high-energy and low energy
parts of the spectrum increases with increasing $\rho$.

  ({\it iv}) The part of the spectrum corresponding to
$xy<4$ is relatively enhanced, since there is no killing
process with these $xy$.

\section{Optimization}\label{secopt}

At $x<4.8$, an increase in the optical length of the
photonic target leads to a monotonic (but limited) rise
in the number of photons (with a simultaneous increase
in the background).
At $x>4.8$, the killing process
stops this rise, and for very large $z$, it kills almost all
high-energy photons.

The dependence of the number
of photons $n(y,z)$ on $z$  has a maximum for some $z=z_m(y)$.
This can be interpreted as the optimal value of $Z$.
There is a question: what criterion should be used
for the optimal choice?

The simplest approach is to consider this balance only for photons of maximal energy, at $y=y_M$ \cite{Tel2001}.
We find it more reasonable to consider for this goal  the $z$-dependence of
entire luminosity  within its high energy peak ${\cal L}_{h.e.}$ \eqref{Lumintdef}, \eqref{taudef}, Table~\ref{hepeak}.

({\it At $\Lambda_C\approx -1$,  the photon energy spectra are concentrated in the narrow band near  $y=y_M$. Therefore, the results of both optimizations are close to each other.
At $\Lambda_C=1$ the initial spectra are flat, and  the estimates
 made for  $y=y_M$ \cite{Tel2001} give an unsatisfactory description of the luminosity}.)\label{pagebad}

\begin{wrapfigure}[12]{l}[0pt]{6.5cm}
\includegraphics[width=0.45\textwidth,height=0.15\textheight]{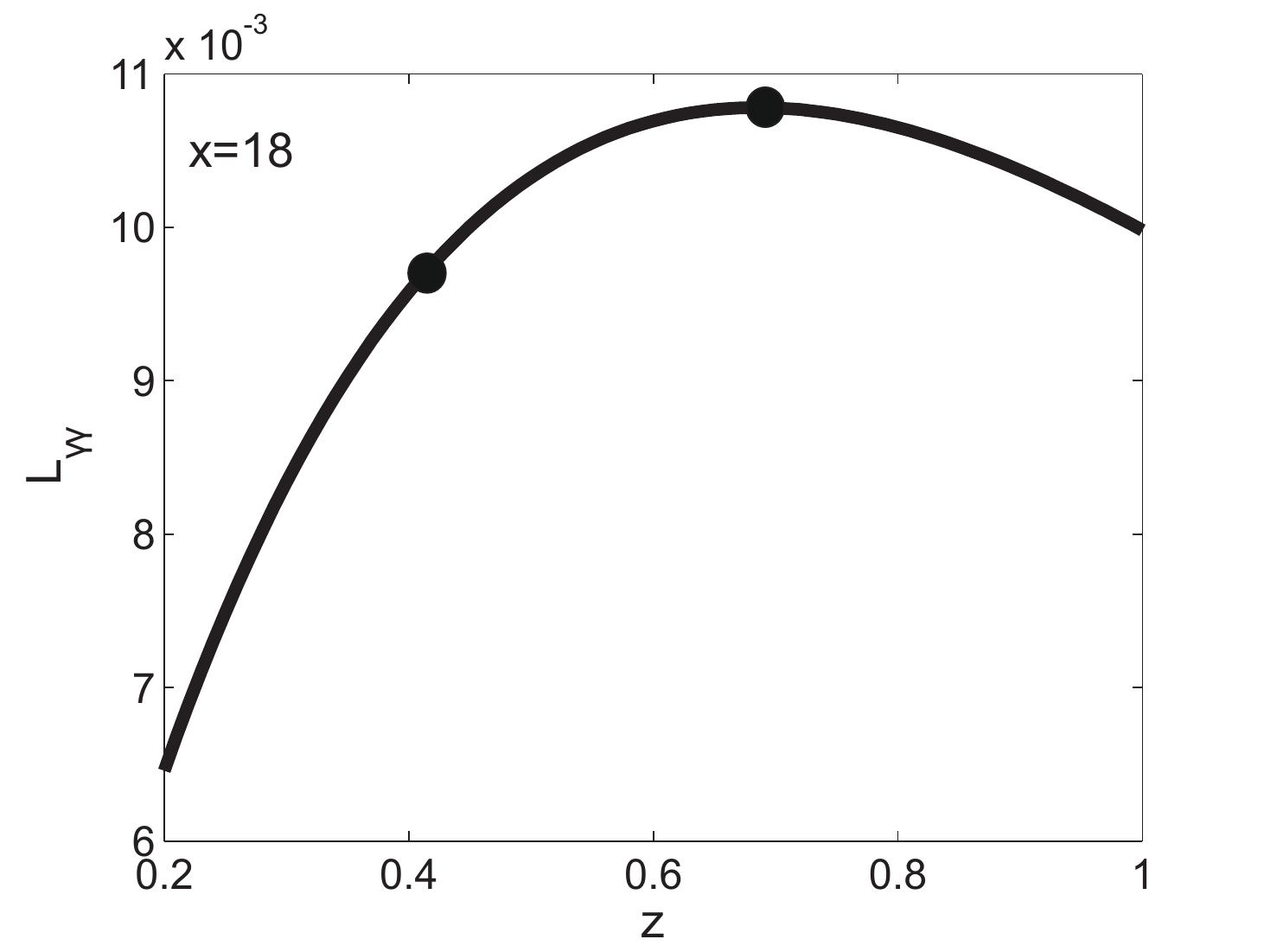}
\vspace{-6mm}
\caption{\it The \ggam \ luminosity integral ${\cal L}_{h.e.}$ in dependence on $z$,
for $\rho=1$, $\Lambda_C=-1$. Dots correspond $z_m$ and $z_{0.9}$.} \label{figlumz}
\end{wrapfigure}
A  typical dependence  of  luminosity ${\cal L}_{h.e.}$ on $z$  is shown in   Fig.~\ref{figlumz}.
The curves at another $x$ and $\rho$ have similar form.
The
optimal value  of the laser optical length are given by the position of a maximum at these curves, $z_m$.
We found numerically that the  value of $z_m$ is practically independent of $\rho$.
These values at $\Lambda_C=-1$ and different $x$ are given in  table~\ref{1bbb}.  In addition to $z_m$, we include here
the value $z_{0.9}(x)<z_m(x)$, provided luminosity which
is 10\% lower than maximal luminosity. In addition, this Table  contains
the energy of the laser flash $A$ \eqref{ZAcorresp} necessary to obtain these optical lengths (in terms of the energy of the laser flash $A_0$ needed to obtain $z = 1$ for $x = 4,5 $);
 the fraction of photons spent on the production of \epe \ pairs, among  photons with highest energy, generated in the basic Compton effect  $r_K(z, y_M) $ (obtained numerically); fraction of electrons freely passing through  the laser bunch
  $d(z) = e^{- z} $.

\begin{table}[ht]
\begin{center}
 \begin{tabular}{||c|c||c|c|c|c|c|c|c|c|}\hline
x&$y_{\min}$&z&$A(z)/A_0$&$r_K(z,y_M)$&
d(z)\\\hline
9&0.7&$z_m=0.704$&1.15&0.22&0.495\\ \cline{3-6}
&&$z_{0.9}=0.49$&0.8&0.13&0.61\\\hline
18&0.75&$z_m=0.609$&1.70&0.43&0.54 \\ \cline{3-6}
&&$z_{0.9}=0.418$&1.17&0.28&0.66\\\hline
100&0.94&$z_m=0.48$&6.3&&0.62\\\cline{3-6}
&&$z_{0.9}=0.32$&4.2&&0.73\\\hline
\end{tabular}
\caption{\it Numbers related to optimization at  $\Lambda_C = -1 $:}
\label{1bbb}
\end{center}
\end{table}

\section{Luminosity distributions}\label{secdistr}

\begin{figure}[ht]
\includegraphics[width=0.48\textwidth,height=0.2\textheight]{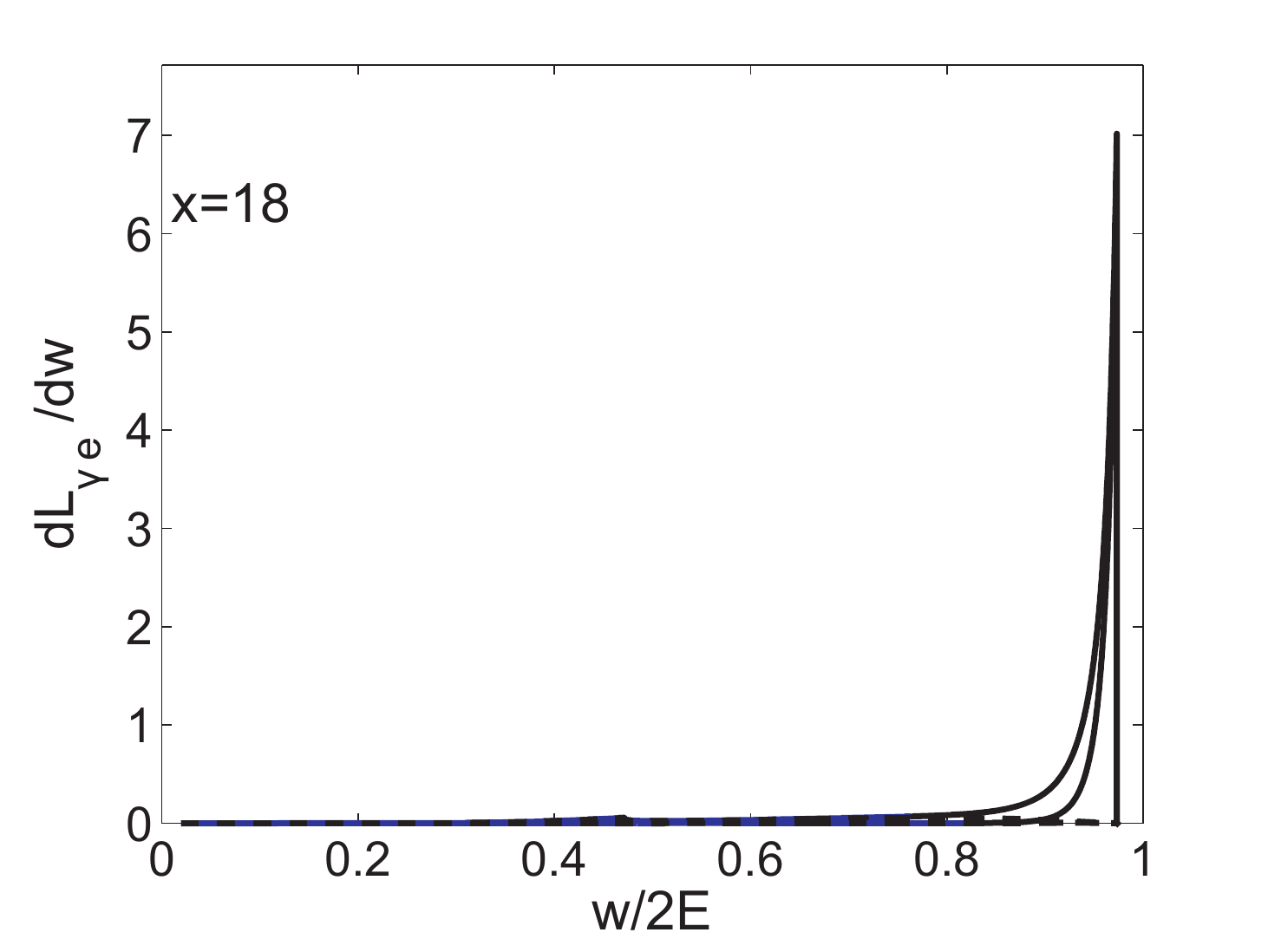}
\includegraphics[width=0.48\textwidth,height=0.2\textheight]{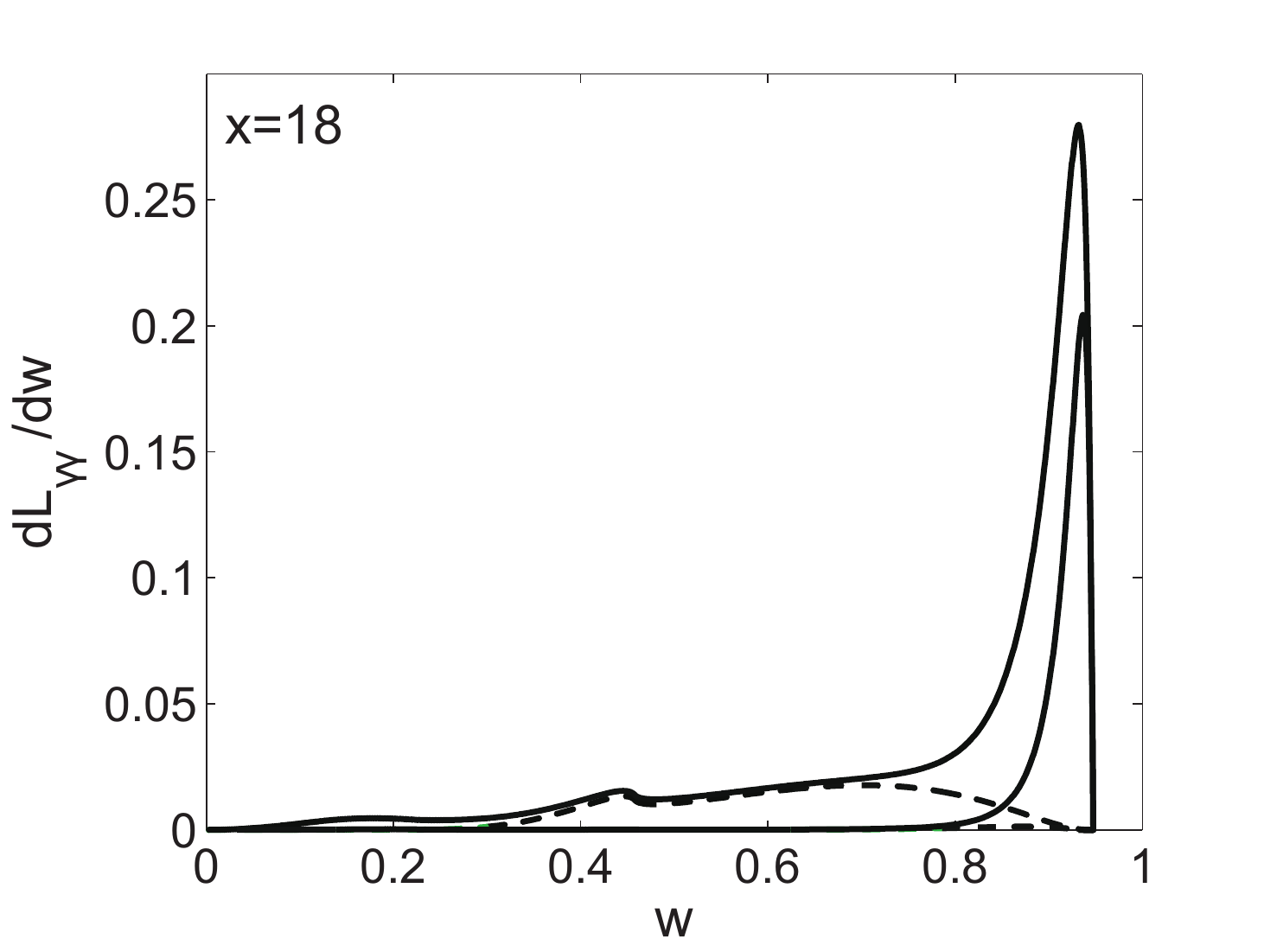}
 \label{lum18}
\caption{\it Luminosity spectra $dL/dw$ at $x=18$, $\Lambda_C=-1$, $z=1$  for $\rho=1$ (upper curves) and $\rho=5$ (lower curves) -- solid lines; dotted lines -- similar distributions for final states with total helicity $3/2$   ($\gamma e$) or 2 (\ggam) at $\rho=1$.}

\end{figure}

Figure.~\ref{lum18}  presents spectra of high energy luminosity for \ggam \ and \egam \ collisions at $x=18$, $\Lambda_C=-1$ for $z=z_m$ $\rho=1$ and 5. (The curves for other $x$ and $\rho$ look similar.) The tables~\ref{tabggamlumLam1}  and  \ref{tabegamlum45918} represent properties of these
luminosity  spectra at $x=9$ and $18$ for $\Lambda_C=-1$ and $\Lambda_C=-0.86$ at $z=z_m$. (The table rows  for $x=4.5$, $z=1$ and $x=100$, $z=z_m$ are presented  for comparison.)

 \begin{table}[ht]
\begin{center} \begin{tabular}{||c|c|c|c|c|c|c|c |c|c|c|c||}\hline
$\rho$& $\Lambda_C$&${\cal L}_{h.e.}$&$L_2/L$&$L_m$&$w_M$&$w_-$&$w_+$&$\gamma_w$ \\\hline
\multicolumn{9}{|c|}{$\ggam: \qquad x=4.5,\qquad z=1,\;\;$}\\\hline
$\ba{c} 1\\ \\5\ea$&$\ba{c}-1\\-0.86\\\hline-1\\-0.86\ea$&
$\ba{c} 0.121\\0.114\\0.031\\0.0275\ea$&$\ba{c} 0.143\\0.215\\0.041\\0.086\ea$&$\ba{c} 0.933\\0.82\\0.464\\0.40\ea$
&$\ba{c}0.779\\0.778\\0.799\\0.7985\ea$&$\ba{c}0.689\\0.672\\0.760\\0.0758\ea$&$\ba{c}0.89
\\0.809\\0.814\\0.813\ea$&$\ba{c}0.154\\0.177\\0.067\\0.069
\ea$\\\hline\hline\hline

\multicolumn{9}{|c|}{$\ggam:\qquad x=9,\qquad z=z_m=0.704,\;\;$}\\\hline

$\ba{c} 1\\ \\5\ea$&$\ba{c}-1\\-0.86\\\hline-1\\-0.86\ea$&$\ba{c}0.0214\\0.0201\\0.0072\\0.064
\ea$&$\ba{c} 0.079\\0.164\\0.021\\0.074\ea$&$\ba{c} 0.222\\0.195\\0.137\\0.118\ea$
&$\ba{c} 0.872\\0.871\\0.885\\0.885\ea$&$\ba{c} 0.814\\0.806\\0.854\\0.852\ea$&$\ba{c} 0.894\\0.894\\0.896\\0.896\ea$&$\ba{c} 0.092\\0.100\\0.048\\0.050\ea$\\\hline\hline\hline

\multicolumn{9}{|c|}{$\ggam: \qquad x=18,\qquad z=z_m=0.609,\;\;$}\\\hline
$\ba{c} 1\\5\ea$&$\ba{c}-1\\-0.86\\\hline-1\\-0.86\ea$&$\ba{c}0.0178\\0.0178\\0.0074\\0.069
\ea$&$\ba{c} 0.089\\0.228\\0.021\\0.144\ea$
&$\ba{c} 0.2615\\0.229\\0.190\\0.164\ea$&$\ba{c} 0.9317\\0.931\\0.9365\\0.936\ea$&$\ba{c} 0.8932\\0.886\\0.9138\\0.912\ea$&$\ba{c} 0.9436\\0.09436\\0.9447\\0.09447\ea$&$\ba{c} 0.054\\0.062\\0.033\\0.035\ea$\\\hline\hline
\hline

\multicolumn{9}{|c|}{$\ggam: \qquad x=100,\qquad z=z_m=0.477,\;\;$}\\\hline
$\ba{c} 1 \\5\ea$&-1&$\ba{c}0.0093\\0.0070
\ea$&$\ba{c} 0.017\\0.009\ea$
&$\ba{c} 0.527\\0.519\ea$
&$\ba{c} 0.9867\\0.9867\ea$&$\ba{c} 0.9771\\0.9793\ea$&$\ba{c} 0.9890\\0.9890\ea$&$\ba{c} 0.012\\0.0097\ea$\\\hline
\hline

\end{tabular}
\caption{\it
Properties of high energy \ggam \ luminosity for different $x$ and $\Lambda_C$.  }
\label{tabggamlumLam1}
\end{center}\end{table}

\begin{table}[ht]\begin{center}
 \begin{tabular}{|c||c|c|c|c||c|c |c|c|}\hline
$\rho$&${\cal L}_{h.e.}$&$L_m$&$L_{3/2}/L$&$\gamma_w$
&${\cal L}_{h.e.}$&$L_m$&$L_{3/2}/L$&$\gamma_w$\\\hline
\multicolumn{9}{|c|}{$x=4.5,\qquad z=1, \qquad w_M=0.9045$}
\\\hline\hline

$\ba{c} 1\\5\ea$&$\ba{c} 0.38\\0.143\ea$&7.626&
$\ba{c} 0.135\\0.007\ea$&$\ba{c} 0.0296\\0.0141\ea$&
$\ba{c} 0.372\\0.134\ea$&7.034&$\ba{c} 0.182\\0.079\ea$&$\ba{c} 0.031\\0.014\ea$\\\hline
&
\multicolumn{4}{|c||}{$\Lambda_C=-1$}&\multicolumn{4}{c|}{$\Lambda_C=-0.86$}\\ \hline\hline\hline

\multicolumn{9}{|c|}{$x=9,\qquad z=0.7047, \qquad w_M=0.949$}
\\\hline\hline
$\ba{c} 1\\5\ea$&$\ba{c} 0.153\\0.074\ea$&4.837&$\ba{c} 0.069\\0.05\ea$&$\ba{c} 0.0176\\0.0105\ea$&
$\ba{c} 0.149\\0.068\ea$&4.32&$\ba{c} 0.116\\ 0.063\ea$&$\ba{c} 0.018\\0.011\ea$\\\hline&
\multicolumn{4}{|c||}{$\Lambda_C=-1$}&\multicolumn{4}{c|}{$\Lambda_C=-0.86$} \\\hline\hline\hline

\multicolumn{9}{|c|}{$x=18,\qquad z=0.609, \qquad w_M=0.973$}
\\\hline\hline

$\ba{c} 1\\5\ea$&$\ba{c} 0.136\\0.08\ea$&7.015&$\ba{c} 0.0625\\0.0059\ea$&$\ba{c} 0.0098\\0.0070\ea$&
$\ba{c} 0.138\\0.076\ea$&6.439&$\ba{c} 0.12\\0.077\ea$&$\ba{c} 0.01\\0.007\ea$\\\hline
&
\multicolumn{4}{|c||}{$\Lambda_C=-1$}&\multicolumn{4}{c|}{$\Lambda_C=-0.86$}\\ \hline\hline\hline

\multicolumn{9}{c|}{$x=100,\qquad z=0.477, \qquad w_M=0.995$}
\\\hline\hline

$\ba{c} 1\\5\ea$&$\ba{c} 0.099\\0.083\ea$&22.905&
$\ba{c} 0.0086\\0.0042\ea$&$\ba{c} 0.002\\0.0018\ea$&
$\ba{c} 0.1085\\0.088\ea$&21.905&$\ba{c} 0.108\\0.103\ea$&$\ba{c} 0.002\\0.0018\ea$\\ \hline
&
\multicolumn{4}{|c||}{$\Lambda_C=-1$}&\multicolumn{4}{c|}{$\Lambda_C=-0.86$}\\ \hline\hline
\end{tabular}
\caption{\it
Properties of high energy \ \  \egam \ luminosity for different $x$ and $\Lambda_C$.  }
\label{tabegamlum45918}
\end{center}
\end{table}

Let us list the important properties of these  distributions for
$x=9$ and $18$:

(1) At the electron beam energy $E=1$~TeV, the maximal photon energy is $\omega_m=0.95$~TeV ($\sqrt{s_{\ggam}}
\approx 1.9$~TeV).

(2)  {\it High-energy luminosity at $\rho=1$}:\\${\cal L} _{h.e.}^{\ggam} \approx 0.02L_{\geom}$ (annual $>
10$ fb$^{-1}$), it is
about $5.5$ times less than that for $x=4.5$;  ${\cal L} _{h.e.}^{\egam} \approx 0.15L_{\geom}$,
about 2.5 times less than that for $x=4.5$.

(3) {\it Peak differential luminosity}\\ $L_m^{\egam} (x=9)\approx L_m^{\egam} (x=4.5)$ and does not depend on $\rho$;
$L_m^{\ggam} (x=9,\rho=1)\approx 0.25L_m^{\ggam} (x=4.5,\rho=1)$.

(4) {\it As $\rho$ increases}, the integrated luminosity
and the peak value decrease, but this decrease is
slower than for $x=4.5$.

(5)  {\it Photons within considered peaks are well polarized}: at $\rho=1$ the fraction of luminosities
${\cal L}_2$ or  ${\cal L}_{3/2}$ in the total luminosities is small; at $\rho=5$ these fractions are negligible\fn{\it A simultaneous change in the signs of helicity of one of the
electrons of LC and a laser photon, colliding with it, lead to substitutions
 $L_0\leftrightarrow L_2$, $L_{1/2}\leftrightarrow L_{3/2}$.}.

(6) {\it The energy distribution of luminosity is very
narrow}: for \ggam \ collisions, the peak width is comparable
to the peak width in the  \epe collision
mode (considering the radiation in the initial state
(ISR) and beam radiation (beamstrahlung, BS)). For \egam \
collisions, the peak width is even narrower than
in the basic \epe collision (considering ISR and BS).
 {\it The  rapidities of produced \egam \ and \ggam \ systems} in the collider rest frame  are contained within a  narrow intervals, determined by the spread of photon energies within high-energy peak \eqref{taudef},
    \beg
    \eta_{e\gamma}\in\left(\fr{1}{2(x+1)},\; \fr{1}{2(x+1)}+\bar{\tau}\right),\quad
    |\eta_{\gamma\gamma}|\le\bar{\tau},\;\;with \;\; \bar{\tau}\approx\fr{\tau}{2}\,.\label{rapidggam}
    \eeg

(7) {\it  Imperfect polarization of the initial electrons} $2\lambda_e=-0.86$ instead of $2\lambda_e=-1$
only weakly degrades the spectra.

\section{Summary }\label{secsum}

(1)  The LC with  electron energy $E \le 1$~TeV allows one
to construct a photon collider (TeV PLC)  using the same lasers and optical systems as those
designed for construction PLC at $E\le 250$~GeV. In comparison
with that case, the required laser flash energy should be increased
 by no more than $70-20$\%.

 The total luminosity integral for  high energy part of spectrum is high enough.
In this part  photon energies are very close to $E$,  photons are monochromatic with good
accuracy in both energy and polarization.

(3) For TeV PLC two  complements compared to the $x<4.8$ case seem to be very desirable:


(a) the
 magnetic deflection of electrons after conversion (see page~\pageref{pagemagn}),

(b) the selection of events with total observed energy of reaction products $\varepsilon>E_{lim}\sim E$ (see page~\pageref{pageselec}).

(4) The low-energy part of the photon spectrum in IR
includes photons from different channels.
It is highly dependent on the details of experimental
facility. In some variants of this facility the corresponding luminosity can be large
\cite{Tel13}. This part of spectrum can be used to study
more traditional problems (for example, see \cite{FCCall}).


\appendix

\section{Some background processes and related issues}

\subsection{A1. "Bad" initial helicity $\Lambda_C\approx 1$}\label{secbad}
At $\Lambda_C=1$
the  photon spectrum in  the main Compton process is much flatter than in the "good"  case $\Lambda_C = -1 $, see Fig.~\ref{fig:Compsp48}. Therefore, in estimates of integrated luminosity \eqref{Lumintdef} one should use lower value $y_{\min}$, e.g. $y_{\min}=0.6$. The optimum optical length in
this case is higher than that for the "good" case $ \Lambda_C = -1 $,
in particular, $z_m(x=18,\,\Lambda_C=1)=0.827$,
$z_m(x=100,\,\Lambda_C=1)=1.07$.
This requires a laser flash energy that is not much higher than that  for the <<good case>>, $A = 1.43A_0$ for $x = 18$ and $A = 5.4A_0$ for $ x = 100$.

The more important difference is the shape of the luminosity spectrum. This spectrum is much flatter than  one shown in Fig.~\ref{lum18}. The position of its
smeared maximum shifts toward much smaller values
of $w$. Here an  one-parametric description of the spectrum at the beam collisions \eqref{rhodef}-\eqref{Lggam} becomes invalid, details of device construction are essential. In this case, the separation between
high-energy and low-energy parts of the luminosity spectrum is practically absent

With the growth of distance $b$ (Fig.~\ref{fig:basschem}) the low-energy part of luminosity disappears,
the residual peak gives much lower integrated luminosity than that at $ \Lambda_C \approx -1 $.

\subsection{A2. The linear polarization of high energy photon}\label{seclinear}

The linear polarization of high energy photon is expressed via
linear polarization of the laser photon $P_{\ell 0}$ by well known ratio $(N/D)$ (see \cite{GKST3}),
in which $|N|\le | 2P_{\ell 0}|$ and $D\propto f(x,y)$.
To reach maximal high energy luminosity in the entire spectrum, the denominator $D$ should be large. Therefore, the linear polarization of photon can be only small in the cases with relatively high \ggam luminosity.
We cannot hope to observe these effects at
the TeV PLCs under discussion.

\subsection{A3. Collisions of positrons with electrons
of the counter-propagating beam}\label{appa}

Collisions of positrons from killing process with electrons of  the opposite beam result in physical states similar to those produced in \ggam collisions. It will be  the main background for  TeV PLC.

{\it General.}
According to Table~\ref{1bbb}, at  $x=18$ and $z=z_m$ the number of killed photons producing high energy \epe \ pairs is less than 3/4 from the number of operative photons. This ratio decreases at lower $x$. Only one half from these photons produces high energy positrons.  Therefore the luminosity of these collisions ${\cal L}_b(\epe)\lesssim (1.5\div 2.5){\cal L}(\ggam)$.

$\lozenge$  The use of the optical length $z_{0.9}$ instead of the optimal one $z_m$  reduces  number of positrons and ${\cal L}_b(\epe)$   by half or even more  with  a small change  in the $L_{\ggam}$.\\

{\it Energy distribution of positrons etc.}  The details of luminosity distribution of these \epe collisions differ
 strongly from those for \ggam collisions of main interest.
To verify this, consider the energy distribution of positrons produced by photons
with energy $\omega=Ey$ and polarization $\lambda$.
  We use notations \eqref{killproc} and denote  the positron energy by $E_+=y_+\omega$. The kinematic constraints are
\bes\label{kinemkill}\beg
w^2=sy\ge 4,\quad v=\sqrt{1-4/w^2}, \quad \fr{1+v}{2}\ge y_+\ge \fr{1-v}{2},
\label{kinemkill1}
\eeg
so
\beg
y_+\le 0.77\to E_+\le 0.693E \;\; \mbox{at}\;\; x=9,\\
y_+\le 0.888\to E_+\le 0.841 \;\; \mbox{at}\;\; x=18\,.\label{killkinem2}
\eeg\ees
As a result,  the rapidity of system produced in these \epe collision is
\be
\eta_{\epe}(x=9)\ge 0.153 ,\quad
\eta_{\epe}(x=18)\ge 0.08. \label{rapidepe}
\ee
These values don't intersect with  the possible rapidity interval for \ggam \ system
\eqref{rapidggam}. Therefore,  the \epe and \ggam \ events are  clearly distinguishable
in the case  of observation of all reaction products.

In general, a more detailed description is desirable.  The energy distribution of positrons produced in the $\gamma_o\gamma$ collision is
\beg
\fr{dn_+(y_+;w^2,\lambda\lambda_o)}{dy_+}
= \fr{(u-2)(1+c\lambda_o\lambda)+s^2}
{\Phi_{\ggam}(w^2,\lambda_o\lambda)};\\
u=\fr{1}{y_+(1-y_+)},\quad c=2\fr{u}{w^2}-1,\quad s^2=1-c^2\,.
\label{posdistrggam}\eeg
For the  highest positron energy  $c=1$, and we have $dn=0$  (at $\lambda_o\lambda=-1$). This
equality $dn=0$  corresponds to the fact that the angular momentum conservation forbids production of positrons (or electrons) in the forward direction. It means that the physical flux of positrons is limited even stronger than that given by Eq.~\eqref{kinemkill}.

Except mentioned endpoints, distribution \eqref{posdistrggam} changes weakly in the whole interval of $y_+$ variation. Therefore, the \epe luminosity is widely distributed over  entire range of its possible variation.  As a result,  the differential luminosity $dL/dw_{\epe}\ll dL/dw_{\ggam}^{peak}$.\\

Apart from the differences in the luminosity distribution,
important {\it differences in the produced systems
at the same energies} should be noted.

{\it (i)} With increasing energy all cross sections in \epe mode  decrease as $1/s$. In the \ggam mode cross sections of many processes don't decrease.

{\it (ii)}  In the \epe mode most of processes are annihilation  ones (via $\gamma$ or $Z$ intermediate states). Products of reaction in such processes have wide angular distribution. In the \ggam mode, a significant
  part of the reaction products move along the collision
axis with a moderate transverse momentum.

\subsection{A4. Bethe-Heitler  process $\boldsymbol{e_o\gamma_o\to ee^+e^-}$}\label{BHproceq}
This process is   switching on at $x=8$. It  is the process of the next order in $\alpha$ but (in contrast to the Compton
effect) its cross section  does not decrease with growth of energy:
\\
\cl{$\sigma_{BH}=(28/9\pi)\alpha \sigma_0 (\ln x-109/42)\;\;$ at $x\gg 1$.}
Because of this process, the yield of high-energy photons decreases by the factor
\be
K_{BH}=\fr{\sigma_C(x)}{\sigma_C(x)+\sigma_{BH}(x)}\,,\label{BHfac}
\ee
and the \ggam \ luminosity reduces  by the factor $K_{BH}^2$.
The numerical values of this factor are presented  in the Table~\ref{tabBH}.
It shows that  the Bethe-Heitler mechanism is negligible at  $x<100$,  the reduction of the photon yield becomes unacceptably large at $x>300$.
\begin{table}[hbt]
\begin{center}
\begin{tabular}{|c||c|c|c|}\hline
$x$ &30&100&300\\ \hline
$\Lambda_C=-1$ &0.96&0.80&0.47\\ \hline
$\Lambda_C=1$ &0.97&  0.88&0.62\\ \hline
\end{tabular}
\caption{\it Factor $K_{BH}$}
\label{tabBH}
\end{center}
\end{table}\vspace{-10mm}

\section{Some physical problems for TeV PLC}\label{secphys}

We expect LHC and $e^+ e^-$ LC to yield  many new results. Certainly, TeV PLC will complement these results and improve precision of some fundamental parameters. However, there are also important problems of fundamental
physics that cannot be studied using the LHC
and \epe \ colliders or require very great efforts to study,
but they can find a solution using TeV PLC. We discuss
just such problems below.

\bu {\bf Beyond Standard Model.}
In the extended Higgs sector one can realizes scenario, in which the observed Higgs boson $h$ is the SM-like (aligned) particle, while model contains  others scalars which interact strongly. It was discussed earlier (for a minimal SM with one Higgs field) that physics of such strongly interacting Higgs sector can be similar to a low-energy pion physics. Such  a system may have resonances like $\sigma$, $\rho$, $f$ with spin 0, 1 and 2 (by estimates, with mass $M\lesssim 1-2$~TeV). High monochromaticity of TeV PLC allows to observe  these  resonance states with spin 0 or 2  in \ggam \ mode.

In the same manner one can observe excited electrons with spin 1/2 or 3/2 in \egam \ mode.

\bu {\bf Gauge boson physics.}

The SM electroweak theory is checked now  at the tree level for simplest processes and at the 1-loop level for $Z$-peak. The
ability to test the effects of this model in more complex
processes and for off- peak loop effects look very
important. A TeV PLC will provide us with a unique
opportunity to explore these problems.

$\triangledown$ Processes \ $\gamma\gamma\to WW$, $e\gamma\to \nu W$ have huge cross sections, by the standards of physics of TeV energies, at large $s$ we have \cite{GKS_W,Jik,Jik1,Ren,Kal}
\be\label{gauge2}
\sigma(\gamma\gamma\to WW)\equiv\! \sigma_V\!=\!8\pi\alpha^2/M_W^2\!\approx\!86~pb, \;
\sigma(e\gamma\to \nu W)\approx\left(\fr{1}{2}-\lambda_e\right) \sigma_V.
\ee\vspace{2mm}

\begin{wrapfigure}[12]{l}[0pt]{5.5cm}\vspace{-2mm}
\includegraphics[width=0.45\textwidth,height=0.19\textheight]{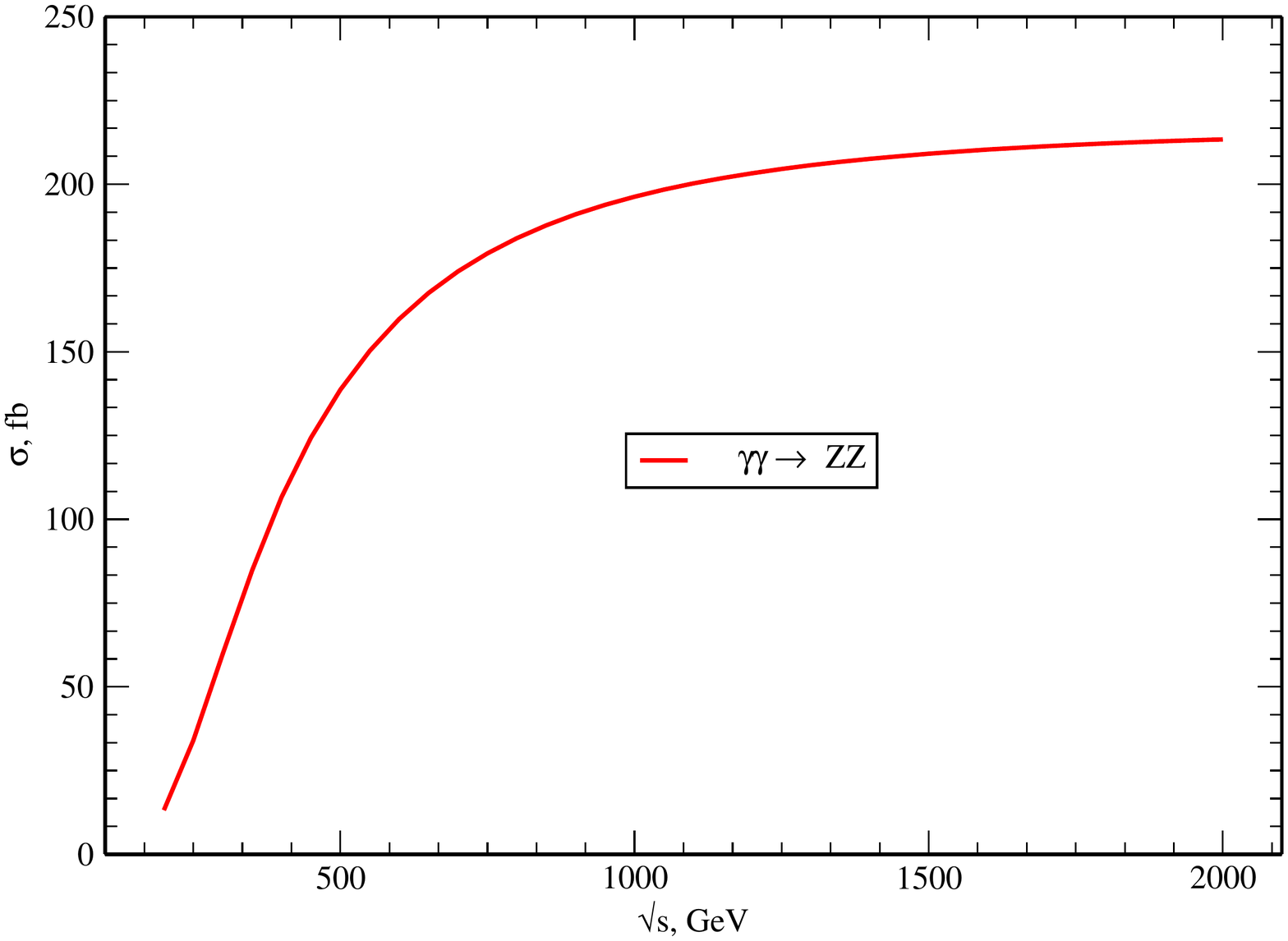}
\vspace{-10mm}
\caption{\it Cross sections  for process $\ggam\to ZZ$ \cite{Kal}.}
  \label{figAAZZ}
\end{wrapfigure}\vspace{-5mm}
 Therefore the measurement  of these processes allows  us  to
test the  detailed structure of the electroweak theory
with  an  accuracy about 0.1\%  (1-loop and partially 2-loop). To
describe the results with such precision, a quantum
field theory with unstable particles should be constructed.

With such precision, the sensitivity to possible anomalous interactions (operators of higher dimension)  --  that is, to the signals of BSM physics --  will be  enhanced \cite{GKS_W}.
The $\ggam\to ZZ$ and $\ggam\to\gamma Z$ processes will be
the first well-measurable processes with variable energy, induced by
only loop contributions \cite{Jik}, \cite{Ren}. The energy dependence for the $\ggam \to ZZ$ cross section  is shown in Fig.~\ref{figAAZZ}. Note that $\sigma(\ggam \to \gamma Z)\approx \sigma(\ggam \to ZZ)/3$. \cite{Kal}.

\begin{figure}[h]
\includegraphics[width=0.48\textwidth,height=0.2\textheight]{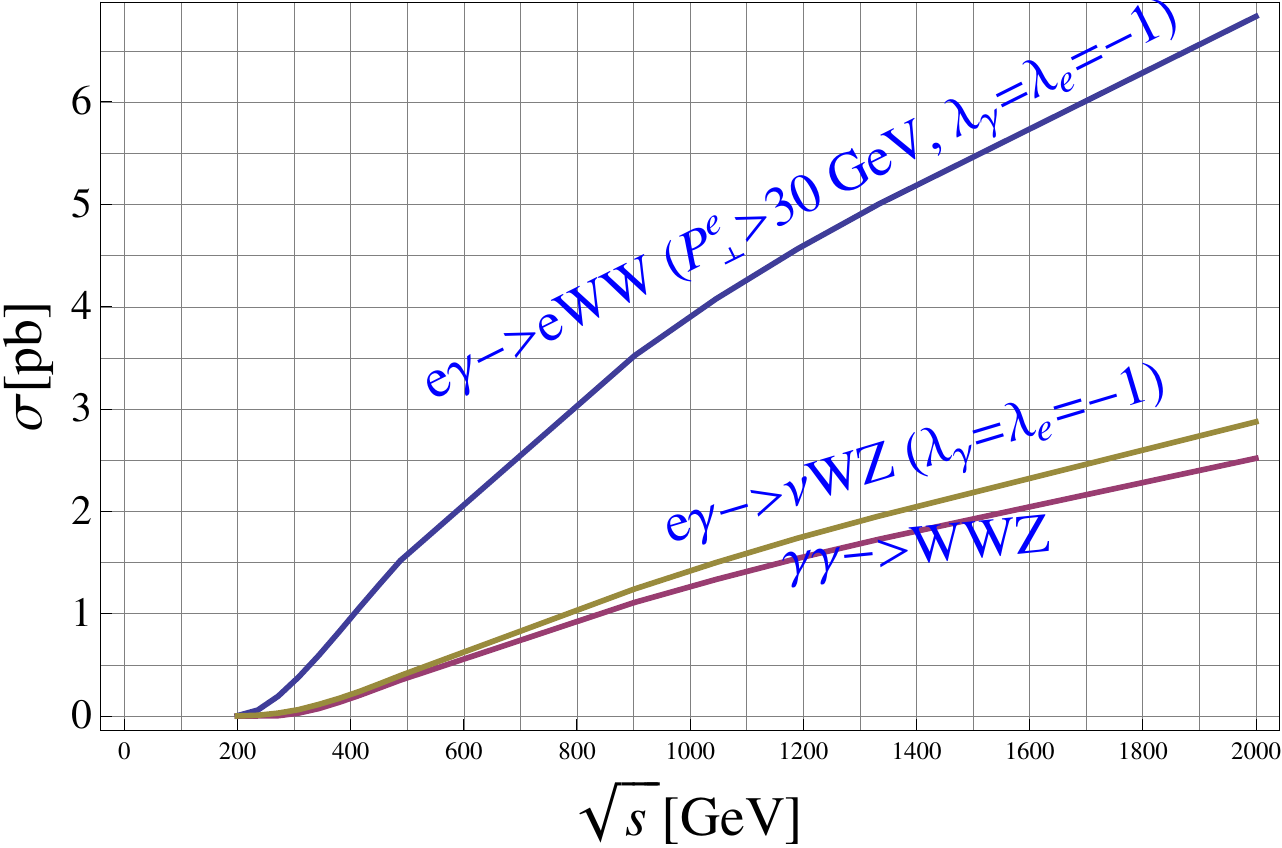}\hspace{5mm}
\includegraphics[width=0.48\textwidth,height=0.2\textheight]{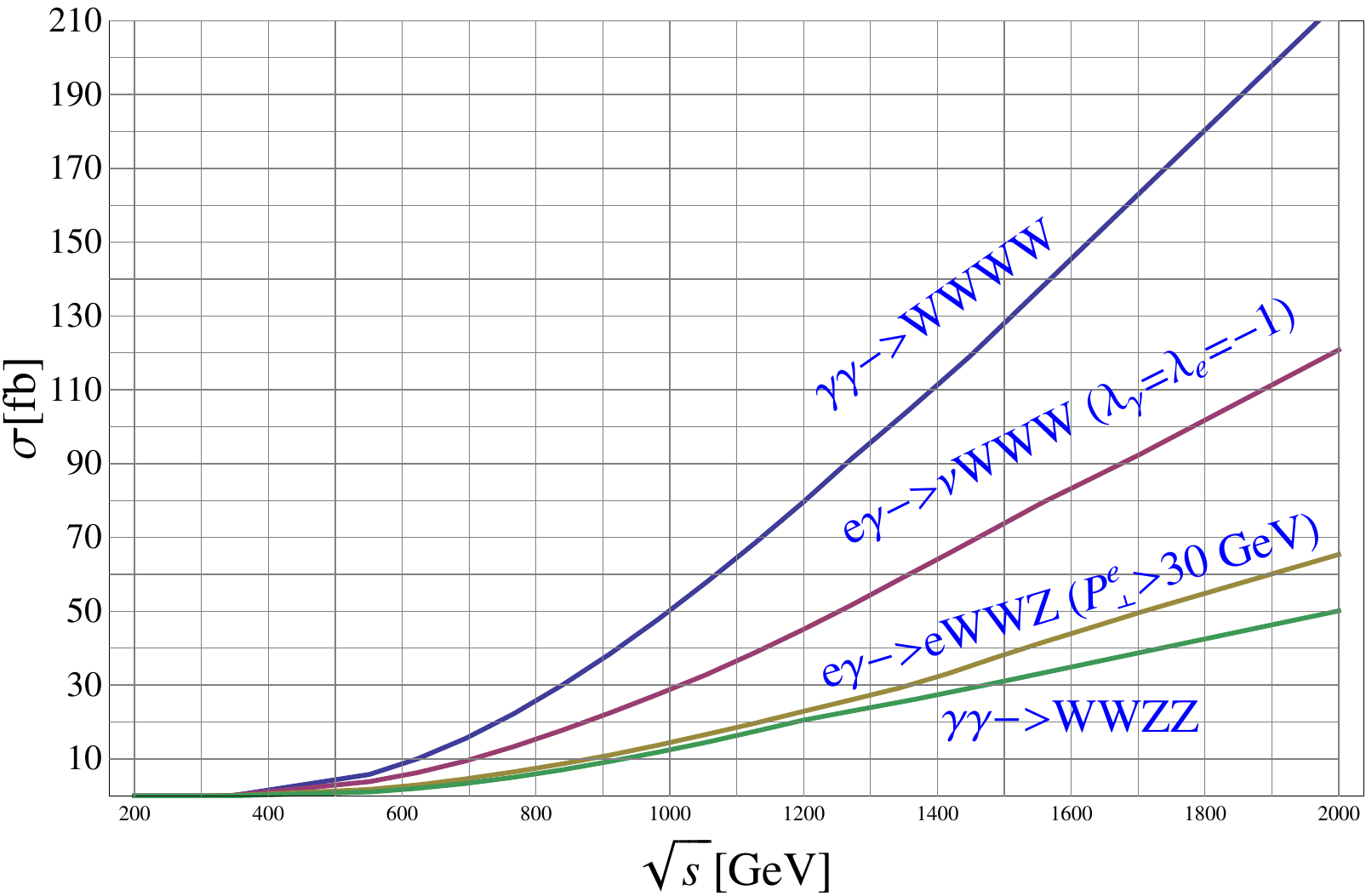}
\caption{\it Cross sections  of multiple production of gauge bosons, calculated with CalcHEP.}
  \label{PROCESS1}
\end{figure}

$\triangledown$ Processes with {\it multiple production of gauge bosons} at TeV PLC have relatively large cross sections, Fig.~\ref{PROCESS1}.
(For $\egam$ collisions we present\lb $\sigma(\egam\to eWW)$, $\sigma(\egam\to eWWZ)$  for well observable transverse momenta of electrons $p_{\bot e}>30$~GeV.)

  The relatively large values of these cross sections are due to the contribution of the diagrams with the exchange of vector bosons in the $t-$ channel, which does not decrease with increasing $s$ \cite{GIS_3}.
They  grow logarithmically with factors $L\approx \ln (s/m_e^2)$ for  photon exchange  or  $\ell\approx \ln(s/M_W^2)$ for  $W(Z)$ exchange:
\beg
\sigma(\egam\to eWW)\sim\alpha\sigma_W L\,,\quad \sigma(\egam\to eWWZ)\sim\alpha^2\sigma_W L\ell\,,\\
\sigma(\egam\to\nu WW)\sim\alpha\sigma_W \ell\,, \quad\sigma(\ggam\to ZWW)\sim\alpha\sigma_W \ell\,,\\
\sigma(\egam\to\nu WWW)\sim\left[\sigma(\ggam\to WWZZ)/\sin^2\theta_W\right]\sim\alpha^2\sigma_W \ell^2\,,
\\ \sigma(\ggam\to WWWW)\sim
\alpha^2\sigma_W \ell^2/\sin^4\theta_W\,.
\eeg

These cross sections are sensitive to the details
of gauge boson interactions (which cannot be seen in another way) and possible anomalous    interactions.  Nothing of the kind can be done on other colliders.
 Studying the dependence of $\egam\to eWW$, $\egam\to eWWZ$ cross sections on the electron transverse momentum will allow to measure the electromagnetic form factor $W$ in the  processes\lb $\gamma\gamma^*\to WW$, $\to WWZ$ and separate the contribution of processes $Z^*\gamma\to WW$, $Z^*\gamma\to WWZ$.
The study of  process $\egam\to \nu WZ$ allows to extract first information about subprocess $W^*\gamma\to WZ$.

\bu {\bf Hadron physics and QCD.}
Our understanding of hadron physics is twofold. We believe that we understand basic theory -- QCD with its asymptotic freedom. However, the  results of calculations in QCD can be applied to the description of data only with the aid of some phenomenological assumptions (often verified by long practice). It results in badly controlled uncertainties in the description of data.

$\triangledown$ PLC is to some extent the hadronic machine with more pure initial state than LHC. Therefore, PLC  can be used also  for detailed study of high energy QCD processes like diffraction, total cross sections, odderon, etc. The results of such experiments can be confronted to  theory with much lower uncertainty than the corresponding ones at the LHC.

$\triangledown$ Studying  the photon structure function $W_\gamma$  (in \egam \ collision) will provide a  unique QCD test. This function can be represented  as the sum of point-like $W_\gamma^{pl}$ and hadronic $W_\gamma^h$ contributions.  The hadronic part is similar to that for proton and is described by a similar phenomenology.
 On the contrary, the point-like contribution $W_\gamma^{pl}$ is described  without phenomenological  parameters \cite{Wit}. The ratio  $W_\gamma^h/W_\gamma^{pl}$ decreases with $Q^2$ roughly as $(\ln Q^2)^{-1/3}$. In the range of parameters accessible today the hadronic contribution dominates.
 The point-like contribution should become dominant with
$Q^2$ growth at TeV PLC.

\section*{Acknowledgments}

We are thankful to   V.~Serbo and V.~Telnov for discussion and comments,   L.  Kalinovskaya and S. Bondarenko -- for information about $\ggam\to ZZ, \,\gamma Z$ processes.
The work was supported by the program of fundamental scientific
 researches of the SB RAS \# II.15.1., project \# 0314-2019-00 and HARMONIA project under contract   UMO-2015/18/M/ST2/00518 (2016-2020).

\end{document}